\documentclass[prl, twocolumn, final, reprint, amsmath, amssymb, superscriptaddress, showpacs, a4paper, aps, 10pt, hidelinks, numbers, sort&compress, floats, balancelastpage]{revtex4-1}

\usepackage{graphicx, color} 
\usepackage[normalem]{ulem}          

\usepackage{hyperref}

\usepackage[utf8]{inputenc}

\usepackage[caption=false]{subfig}
\usepackage{amssymb}
\usepackage{amsmath}
\usepackage{commath}
\usepackage{graphicx,bm}
\usepackage{verbatim}

\newcommand{\1}{\ensuremath{\left|1 \right\rangle}}

\definecolor{britishracinggreen}{rgb}{0.0, 0.26, 0.15}
\definecolor{bulgarianrose}{rgb}{0.28, 0.02, 0.03}
\definecolor{darkred}{rgb}{0.90,0,0}
\definecolor{darkgreen}{rgb}{0,0.60,.2}
\definecolor{darkblue}{rgb}{0,0,1}
\definecolor{orange}{cmyk}{0,0.6,0.8,0}
\definecolor{lightblue}{rgb}{0.3,0.5,1}
\definecolor{lightgreen}{rgb}{0.4,0.80,.4}

\begin{document}

\title{Critical transport and vortex dynamics in a thin atomic Josephson junction}

\author{K. Xhani$^{1,2}$, E. Neri$^{3}$, L. Galantucci$^{1}$, F. Scazza$^{2,4}$, A. Burchianti$^{2,4}$, K.-L. Lee$^1$,\\ C. F. Barenghi$^{1}$, A. Trombettoni$^{6}$, M. Inguscio$^{2,4,5}$, M. Zaccanti$^{2,3,4}$, G. Roati$^{2,4}$ and N. P. Proukakis}

\address{
Joint Quantum Centre (JQC) Durham-Newcastle, School of Mathematics, Statistics and Physics, \\Newcastle University, Newcastle upon Tyne NE1 7RU, United Kingdom\\
\mbox{$^{2}$European Laboratory for Non-Linear Spectroscopy (LENS), Universit\`{a} di Firenze, 50019 Sesto Fiorentino, Italy}\\
$^{3}$Dipartimento di Fisica e Astronomia, Universit\`{a} di Firenze, 50019 Sesto Fiorentino, Italy\\
$^{4}$Istituto Nazionale di Ottica del Consiglio Nazionale delle Ricerche (CNR-INO), 50019 Sesto Fiorentino, Italy\\
$^{5}$Department of Engineering, Campus Bio-Medico University of Rome, 00128 Rome, Italy\\
$^{6}$Istituto Officina dei Materiali del Consiglio Nazionale delle Ricerche (CNR-IOM) and Scuola Internazionale Superiore di Studi Avanzati (SISSA), Trieste, Italy}

\date{\today}



\begin{abstract}

We study the onset of dissipation in an atomic Josephson junction between Fermi superfluids in the molecular Bose-Einstein condensation limit of strong attraction. Our simulations identify the critical population imbalance and the maximum Josephson current delimiting dissipationless and dissipative transport, in quantitative agreement with recent experiments. We unambiguously link dissipation to vortex ring nucleation and dynamics, demonstrating that quantum phase slips are responsible for the observed resistive current. Our work directly connects microscopic features with macroscopic dissipative transport, providing a comprehensive description of vortex ring dynamics in three-dimensional inhomogeneous constricted superfluids at zero and finite temperatures.

\end{abstract}

\maketitle

Interest is growing in model systems that allow for investigating the interplay between resistive and dissipationless quantum transport phenomena. In this context, ultracold gases in tailored optical potentials represent an ideal framework, owing to the real-time control over the relevant parameters in experiments \cite{Ventra, brantut}, combined with the ability for \textit{ab initio} modelling \cite{nick_book, Berloff}.
 A paradigmatic example is the study of the dynamics between two atomic superfluids weakly coupled through a thin tunnelling barrier. This realizes a Josephson junction \cite{JOS, barone}, which represents a minimal platform to observe both coherent quantum transport \cite{Tinkham, barone}, and its breakdown driven by dissipative microscopic mechanisms \cite{caldeira, fazio}.

The coherent dynamics of atomic Josephson junctions (JJs) \cite{MQST1,MQST2,MQST2b,MQST3,JO0,JOTROMBETTONI,JO1,JO2,JO5,JO3,Ic2} 
is  governed by the competition between the charging energy $E_C$  and the Josephson tunneling energy $E_J$ \cite{MQST1,MQST2}. 
 $E_C$ relates the chemical potential difference across the tunnelling barrier to the relative population imbalance between the reservoirs, and depends on interparticle interactions. 
 $E_J$ promotes the delocalization of the superfluid across the two reservoirs and sets the maximum coherent flow through the weak link. When $E_J$ dominates, superfluid current and relative phase oscillate in quadrature at the Josephson plasma frequency. In the opposite regime, and in the absence of dissipation \cite{MQST2,MQST2b}, the system may enter the Macroscopic Quantum Self-Trapping (MQST) regime. This is characterized  by high-frequency coherent oscillations of the population imbalance around a non-zero value, driven by a monotonically increasing relative phase \cite{MQST1,MQST3,JOTROMBETTONI,JO1,JO2,JO5}. Even without thermally induced decay of the population imbalance \cite{JO2, MQST2b, MQST4}, the stability of MQST depends 
on whether vortices nucleated inside the barrier annihilate therein \cite{Piazza1, Abad}, or penetrate into the superfluid reservoirs. Recent experiments with inhomogeneous three-dimensional 
Fermi superfluids \cite{Liscience, Lidiss} revealed the intimate connection between phase slippage and dissipation arising from vortices created within the barrier and shed into the superfluid. Similar effects have been studied in ring-shaped bosonic 
condensates \cite{Wright2013, Jendrzejewski, Mathey2014, Snizhko2016}, mesoscopic structures \cite{Eckel, Gauthier} and lower-dimensional geometries \cite{Derrico2018, Polo2018, interaction}. While vortices crossing the weak link are known to yield a finite resistance \cite{Jendrzejewski, Eckel, Lidiss}, the relation between microscopic vortex nucleation, dynamics and macroscopic dissipative flow is still poorly understood.

 In this work we demonstrate the connection between resistive superfluid currents and vortex ring (VR) dynamics in an atomic JJ of fermionic superfluids. We obtain the critical population imbalance and the maximum coherent current delimiting the boundary between dissipationless and dissipative transport even at finite temperatures. We find excellent agreement with recent measurements \cite{Liscience,Lidiss}, thus clarifying their interpretation. Trap asymmetry is shown to foster the emergence of elliptical VRs exhibiting Kelvin wave excitations, while thermal fluctuations reduce the VR lifetime. 

{\em Methodology.} Our numerical simulations are based on the experimental parameters of Ref.~\cite{Lidiss}.  We consider two molecular Bose-Einstein condensates (BEC) of about  $10^5$ atom pairs of $^6$Li, weakly coupled through a thin optical barrier, at $1/k_Fa \sim 4.6$  (where $k_F$ is the Fermi wavevector and $a$ the interatomic scattering length).  The harmonic trapping potential is asymmetric with approximately (1:12:10) ratio along the $x$, $y$ and $z$-axis, respectively.
The Gaussian barrier bisects the gas along the weakest ($x$) direction (with  $\nu_x =15$ Hz), with a $1/e^2$ waist, $w \sim 4 \xi$, where $\xi$ is the superfluid coherence length 
 \cite{Lidiss}. The superfluid transport through the barrier is triggered by an initial non-zero population imbalance $z_0 = z_\mathrm{BEC}(0)$ between the two reservoirs. Here, $z_\mathrm{BEC}(t)=(N_R(t)-N_L(t))/N_\mathrm{BEC}$, with $N_L\,(N_R)$ the BEC number in the left (right) reservoir, and $N_\mathrm{BEC}=N_L+N_R$ the total condensate number. The imbalance corresponds to a chemical potential difference $\Delta\mu=\mu_L-\mu_R = E_C z_0 N_{\rm BEC}/2$. Dynamics in the $T=0$ limit are simulated via the time-dependent Gross-Pitaevskii equation (GPE), extended to non-zero temperatures $T \lesssim 0.4\,T_c$ (where $T_c$ is the BEC critical temperature), via its coupling to a collisionless Boltzmann equation \cite{nick_book,ZNG2,SM}.  We stress that the standard two-mode model \cite{MQST1,MQST3} that captures both Josephson and MQST dynamics of previous experiments \cite{JO5, JO1} is out of its validity range due to the considered values of the ratio $V_0/\mu$ and to the thinness of the junctions \cite{SM}. Although dissipative effects can be {\em phenomenologically} modeled by damped two-mode \cite{marino1999,MQST4,pigneur2018} and RSJ-circuital models \cite{Tinkham}, such approaches provide limited insight into the microscopic dissipative processes.

\begin{figure}[t!]
\begin{center}
\includegraphics[width= \columnwidth]{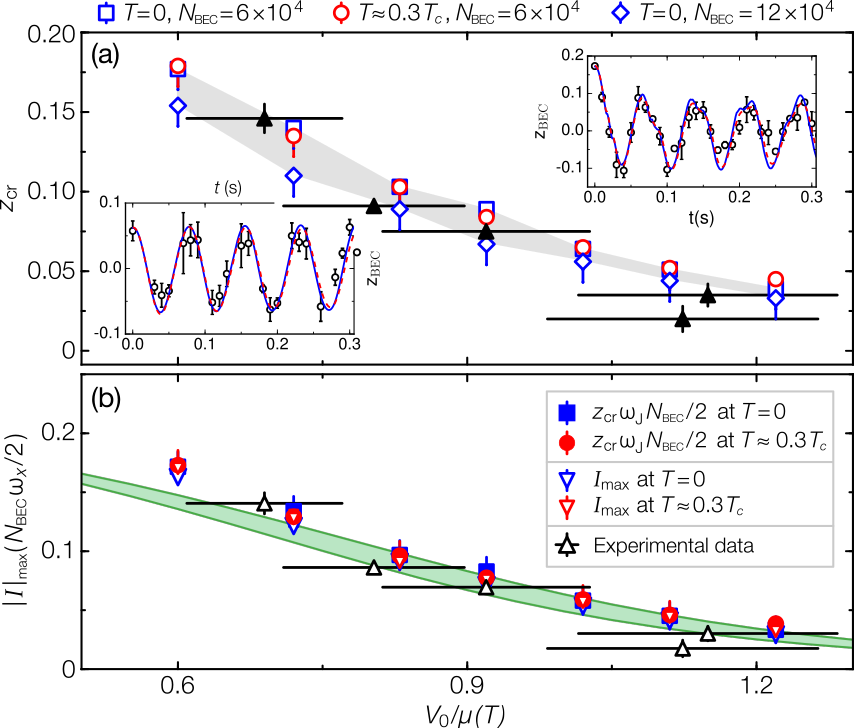}
\vspace{-0.6cm}
\caption{Phase diagram of a thin Josephson junction: 
(a) Critical population imbalance $z_\mathrm{cr}$
as a function of $V_0/\mu(T)$, via numerical simulations at $T=0$ (blue symbols) and $T \approx 0.3 T_c$ (red circles), and experimental data (black triangles). 
Grey shaded area accounts for the experimental range of particle number. 
Vertical error bars are set by the discreteness of the numerically-probed $z_0$ values (simulations) and the standard deviations over at least four measurements (experiments); horizontal experimental error bars are set by the combined uncertainties in measuring barrier width, particle number and laser power. Insets: comparison of numerical (blue and red lines) and experimental population imbalance evolution in Josephson (left) and dissipative (right) regimes.
(b) Maximum superfluid current, $\abs{I}_\mathrm{max}$, based on the numerical time derivative of population imbalance (down triangles), and on the numerical/experimental estimate
$\abs{I}_\mathrm{max} \simeq z_\mathrm{cr} \, \omega _J \, N_\mathrm{BEC}/2$.
Green shaded area: predicted maximum supercurrent (including second-order harmonics in the current-phase relation), accounting for the uncertainty in $V_0/\mu$ \cite{SM}. }
\label{Fig1}
\vspace{-0.4cm}
\end{center}
\end{figure}

{\em Dynamical Regimes and Phase Diagram.} We study $z_\mathrm{BEC}(t)$, varying both the initial population imbalance $z_0$ and the barrier height $V_0$. At each value of $V_0$ we observe two distinct dynamical regimes. For $z_0$ smaller than a critical value $z_\mathrm{cr}$, $z_\mathrm{BEC}(t)$ exhibits sinusoidal plasma oscillations (Josephson regime). For $z_0\geqslant z_\mathrm{cr}$, we instead observe an initial rapid decay of $z_\mathrm{BEC}(t)$ (dissipative regime), followed by plasma oscillations with amplitude smaller than $z_\mathrm{cr}$. We validate our numerics by comparing $z_\mathrm{BEC}(t)$ with experiments under the same conditions, finding excellent agreement [Fig.~\ref{Fig1}(a), insets].
Combining calculated and newly extracted experimental $z_\mathrm{cr}$ values, we map out the phase diagram delimiting Josephson and dissipative regimes as a function of the normalized barrier height $V_0/\mu(T)$ [Fig.~\ref{Fig1}(a)], where $\mu(T)$ is the chemical potential including the thermal mean-field contribution \cite{ZNG2, SM}. Increasing $V_0/\mu(T)$, the onset of dissipation appears at smaller $z_\mathrm{cr}$. This 
reproduces the
observed boundary within experimental uncertainty up to $T\approx0.3\,T_c$ upon keeping the {\em condensate} number equal to the $T=0$ case.
Our findings can also be interpreted in terms of the critical current $I_\mathrm{max}$ across the junction, defined as the maximum value of $I=\dot{z}_\mathrm{BEC} \,N_\mathrm{BEC}/2$ at $z_0=z_\mathrm{cr}$ [Fig.~\ref{Fig1}(b)]. 
Numerically, $\abs{I}_\mathrm{max}$ is well approximated by $z_\mathrm{cr} \, \omega _J \, N_\mathrm{BEC}/2$, where $\omega _J$ is the Josephson plasma frequency. The corresponding $\abs{I}_\mathrm{max}$ from the experimentally determined $z_\mathrm{cr}$ and $\omega _J$ is in excellent agreement with the theoretical prediction. The overall trend of $\abs{I}_\mathrm{max}$ against $V_0/\mu(T)$ is also quantitatively captured by extending to inhomogeneous systems an analytical model, originally developed for two homogeneous BECs weakly coupled through a rectangular barrier \cite{Ic2,SM, zaccanti}.
\begin{figure}[t!]
\begin{center}
\includegraphics[width= \columnwidth]{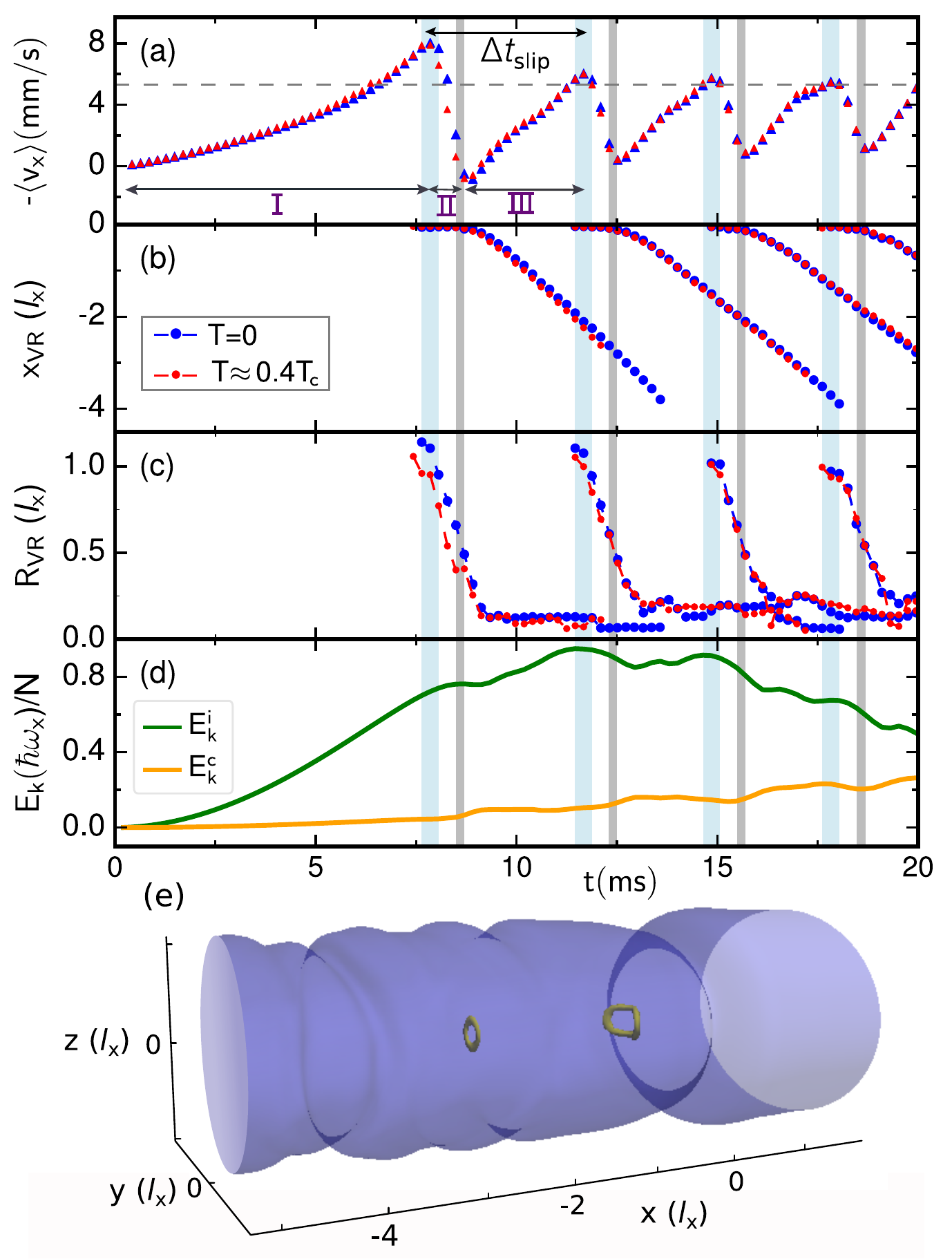}
\vspace{-0.5cm}
\caption{Vortex ring generation and early-stage dynamics.
(a) Density-weighted $x$-component of superfluid velocity at the barrier (dashed line denotes mean speed of sound $\langle c \rangle=\sqrt{\mu/2M}$). 
(b) Mean radius and (c) position of the first few generated vortex rings (units of $l_x=\sqrt{\hbar/ M \omega _x}$). (d) Temporal evolution of the  incompressible ($E_k^i$) and compressible ($E_k^c$) kinetic energy of the BEC. 
Vertical shaded blue areas denote the maxima of superfluid velocity when VRs are generated, while grey areas indicate the times when the VRs enter the Thomas-Fermi surface. 
Shown are both $T=0$ (blue symbols) and $T\approx0.4\,T_c$ (red symbols) for $z_0=0.25$, $V_0/\mu\simeq 0.8$ and $N_\mathrm{BEC} \simeq 6 \times10^4$. (e) BEC density isosurface at 19.5\,ms, when the  $3^{\rm rd}$ and  $4^{\rm th}$ VRs are visible.
}
\vspace{-0.4cm}
\label{Fig2}
\end{center}
\end{figure}

{\em Vortex Ring Nucleation and Evolution.}  The onset of the dissipative regime for $z_0 \geq z_\mathrm{cr}$ in \cite{Liscience,Lidiss} has been linked to the appearance of topological defects in the superfluid. Here, we fully characterize such dynamical features at $T = 0$, by computing the superfluid velocity $\mathbf{v}=(\hbar/M) \nabla \phi$, where $M$ is the atom pair mass and $\phi$ the condensate phase 
(see later for thermal effects).
Given the symmetry of our junction, we consider the $x$-component of the superfluid velocity, weighted over the transverse density in the $x=0$ plane, $\langle v_x \rangle$ \cite{SM}; 
we identify three distinct dynamical stages (I, II and III, see Fig.~\ref{Fig2}) in the nucleation process of the first VR (the emerging pattern applies to subsequent VRs). 
In stage I, following the Josephson-Anderson relation $M \dot{\bf{v}} = -\nabla \mu$ \cite{Anderson,Varoq,sato-packard-2012}, the chemical potential gradient $\nabla \mu$ drives an unidirectional, accelerated superfluid flow across the junction. When $-\langle v_x \rangle$ reaches a critical value 
exceeding the mean 
sound speed $\langle c \rangle$ (Fig.~2(a)), a VR is nucleated, associated with a relative-phase jump of $2 \pi$ \cite{SM}. It originates outside the Thomas-Fermi (TF) surface on the central radial plane, $x=0$, where, the superfluid velocity is maximum, due to the flow constriction, and the local speed of sound is minimum, since density vanishes. 
After its nucleation [stage II], the VR moves axially very slowly away from its nucleation region $x_\mathrm{V\!R}=0$ 
(Fig.~\ref{Fig2}(b)) with its mean radius $R_\mathrm{V\!R}$ rapidly 
shrinking (Fig.~\ref{Fig2}(c)) 
due to the strong radial density inhomogeneity in the barrier region,
until it is comparable to the transversal TF radius  of the BEC, and enters the bulk superfluid \cite{Piazza1}. 
During such evolution, $-\langle v_x \rangle$ exhibits a rapid decrease, possibly even changing sign. 
Then, in stage III, the VR gradually leaves the barrier region with the axial velocity  $-\langle v_x \rangle$
re-accelerated by $\nabla \mu$ (Fig.~\ref{Fig2}(a)), until some time later ($\Delta t_\mathrm{slip}\simeq h/\Delta \mu$), when it has already travelled a considerable distance from the barrier edge, another VR is nucleated at the trap centre  (see, e.g.~Fig.~\ref{Fig2}(e)). 
Note that early on in stage III, before the VR exits the barrier region (i.e.~before reaching the point of maximum transversal TF radius),  $R_\mathrm{V\!R}$ continues decreasing due to the strong background density gradient.
%


%



For a deeper insight into the underlying superfluid dynamics, we decompose at $\bf{x=0}$ the total axial superfluid velocity $v_x=v_f+v_\omega$, where $v_f$ is
the main flow velocity (which is slowly varying compared to the timescale of the early vortex dynamics) and $v_\omega$ is the VR-generated swirling velocity \cite{SM}; we also initially neglect compressibility effects (addressed in the next paragraph).
By the end of stage II, the shrinking VR has just left the trap centre, and so the vortex contribution  $v_\omega$ evaluated at $\bf{x=0}$ (where the local superfluid velocity $\langle v_x \rangle$ shown in Fig.~\ref{Fig2}(a) is calculated) tends to 0.
%
This leads to a drop of $\langle v_x \rangle$ with amplitude 
$\Delta\langle v_x \rangle \sim \kappa/R_\mathrm{V\!R}$, corresponding to the change in the axial superfluid velocity at the trap centre due to the lost vortex contribution, where $\kappa$ is the quantum of circulation \cite{SM}.
This sawtooth-like profile of $\langle v_x \rangle$
(Fig.~\ref{Fig2}(a)) is typical of phase slippage phenomena seen in superfluid helium \cite{Anderson, Varoq, hoskinson-2006, sato-packard-2012}, with the less abrupt drop found here stemming from the initial persistence of the VR within the barrier region.

The drop $\Delta\langle v_x \rangle$ can even overcome the generating flow velocity, leading to flow reversal (i.e.~backflow) in the post-nucleation dynamics, in agreement with Biot-Savart calculations \cite{SM}. 
The amplitude of each subsequent velocity drop is reduced due to the overall decay of $z_\mathrm{BEC}(t)$.

To connect the dissipation with the microscopic VR nucleation and dynamics, and phonon emission, we decompose the temporal evolution of the BEC total kinetic energy in its incompressible $E_k^i$, compressible $E_k^c$, and quantum pressure $E_q$ contributions \cite{nore1997,SM}. $E_k^i$ and $E_k^c$ correspond respectively to the kinetic energy of the flow (both potential flow driven by $\nabla \mu$ and vortex generated swirls) and to the sound wave energy in the superflow. $E_q$ 
accounts for the energy arising from density inhomogeneities due to the trapping potential
\cite{SM}. 
When each VR enters the TF surface 
(end of stage II), and while still propagating within the barrier's region of increasing density,
sound waves are emitted and $E_k^c$ increases at the expenses of $E_k^i$ (Fig.~\ref{Fig2}(d)) \cite{SM}.
The dissipation of Josephson oscillations \cite{Anderson, Varoq, sato-packard-2012, Lidiss} thus stems from two effects: the incompressible kinetic energy transferred from the axial flow to the 
vortex swirling flow and the phonon-emission occurring during vortex nucleation and propagation within the barrier region.

\begin{figure}[t!]
\begin{center}
\includegraphics[width= \columnwidth]{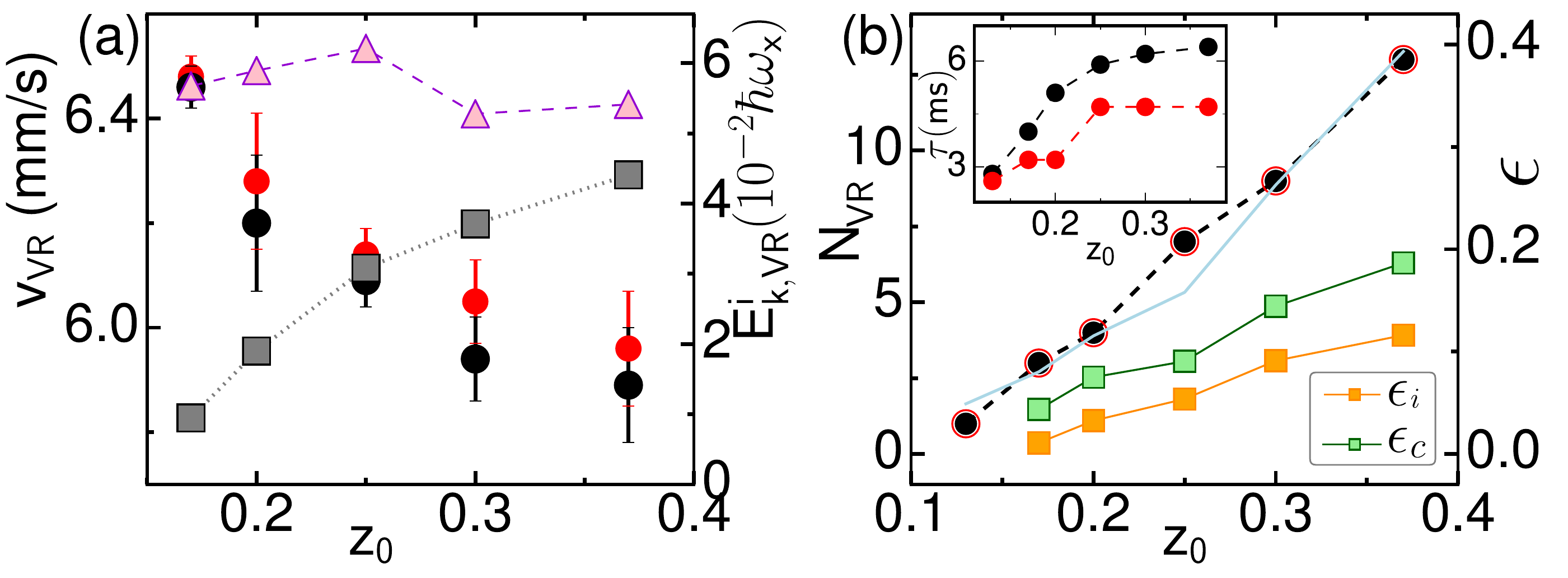}
\vspace{-0.3cm}
\caption{
Role of initial population imbalance $z_0$ (for fixed $V_0/\mu\simeq 0.8$ for which $z_{cr}\simeq 0.11$) on: (a) velocity $v_\mathrm{V\!R}$ (left axis, red/black circles) and incompressible kinetic energy $E_{k,VR}^i$ (right axis, grey squares) of $1^{\rm st}$ nucleated VRs. Pink triangles indicate the VR energy calculated with the analytical formula for homogeneous unbounded BECs \cite{SM}; 
(b) total number $N_{V\!R}$ of VRs penetrating the bulk (left axis, circles) and  
vortex induced dissipations $\epsilon_i$ and $\epsilon_c$ (right axis, yellow/green squares). Blue line connects $N_\mathrm{V\!R}$ estimates from the time-averaged phase-slippage rate 
$\Delta \mu (t)/h$ \cite{Anderson}. 
Inset: lifetimes, $\tau$, of $1^{\rm st}$ nucleated VRs.
Each subplot shows $T=0$ (black symbols) and $T\approx0.4\,T_c$ (red symbols) results.}  
\label{Fig3}
\end{center}
\end{figure}

We further quantify both those effects by considering the effect of $z_0$ on the velocity $v_\mathrm{V\!R}$ and the incompressible kinetic energy $E_{k,V\!R}^i$ of the first VR nucleated; this is shown in  Fig.~\ref{Fig3} for $z_0 \in [0.13,\,0.37]$ and $V_0/\mu=0.8$.
%
We find that increasing $z_0$ leads to a decreasing $v_\mathrm{V\!R}$ and to a monotonic increase of $E_{k,V\!R}^i$ [Fig.~\ref{Fig3}(a)] \cite{SM}. Calculating the fraction of the total kinetic energy flowing through the junction until the nucleation of the first VR which is dissipated in sound ($\epsilon_c$) or transferred to the first VR ($\epsilon_i$)
\cite{SM},
we find that both sources of dissipation increase as $z_0$ gets larger [see Fig.~\ref{Fig3}(b)], and can cumulatively account for a significant fraction of the total energy. Surprisingly, the acoustic dissipation $\epsilon_c$ is always larger than the incompressible contribution $\epsilon_i$.
Increasing $z_0$ leads to more nucleated vortices $N_\mathrm{V\!R}$ (Fig.~\ref{Fig3}(b)), due to the larger time-averaged chemical potential difference \cite{Anderson},  consistently with \cite{Lidiss} and with earlier studies of controlled vortex generation \cite{neely2010, kwon2015}. Similarly, the VR lifetime increases by increasing $z_0$ [Fig.~\ref{Fig3}(b) (inset)]. 
The VR survival during its propagation in the superfluid bulk is thus determined by two competing effects: On the one hand, the VR tends to expand \cite{wang-etal-2017} to conserve its incompressible kinetic energy as it moves towards lower-density regions with decreasing transverse size. 
On the other hand, the radial trapping asymmetry ($\omega_y \neq \omega_z$) leads to elliptical VR profiles with oscillating aspect ratio, corresponding to a $m=2$ Kelvin wave excitation on a circular VR \cite{barenghi-etal-1985}. This wobbling motion induces dissipation of the VR incompressible kinetic energy via the emission of phonon-like excitations, reducing the VR radius \cite{klein2014,galantucci-etal-2019}: When $R_\mathrm{V\!R} \sim \xi$, the VR loses its circulation and annihilates in a rarefaction pulse \cite{Piazza1}. 
\begin{figure}[t!]
\begin{center}
\includegraphics[width=0.99\columnwidth]{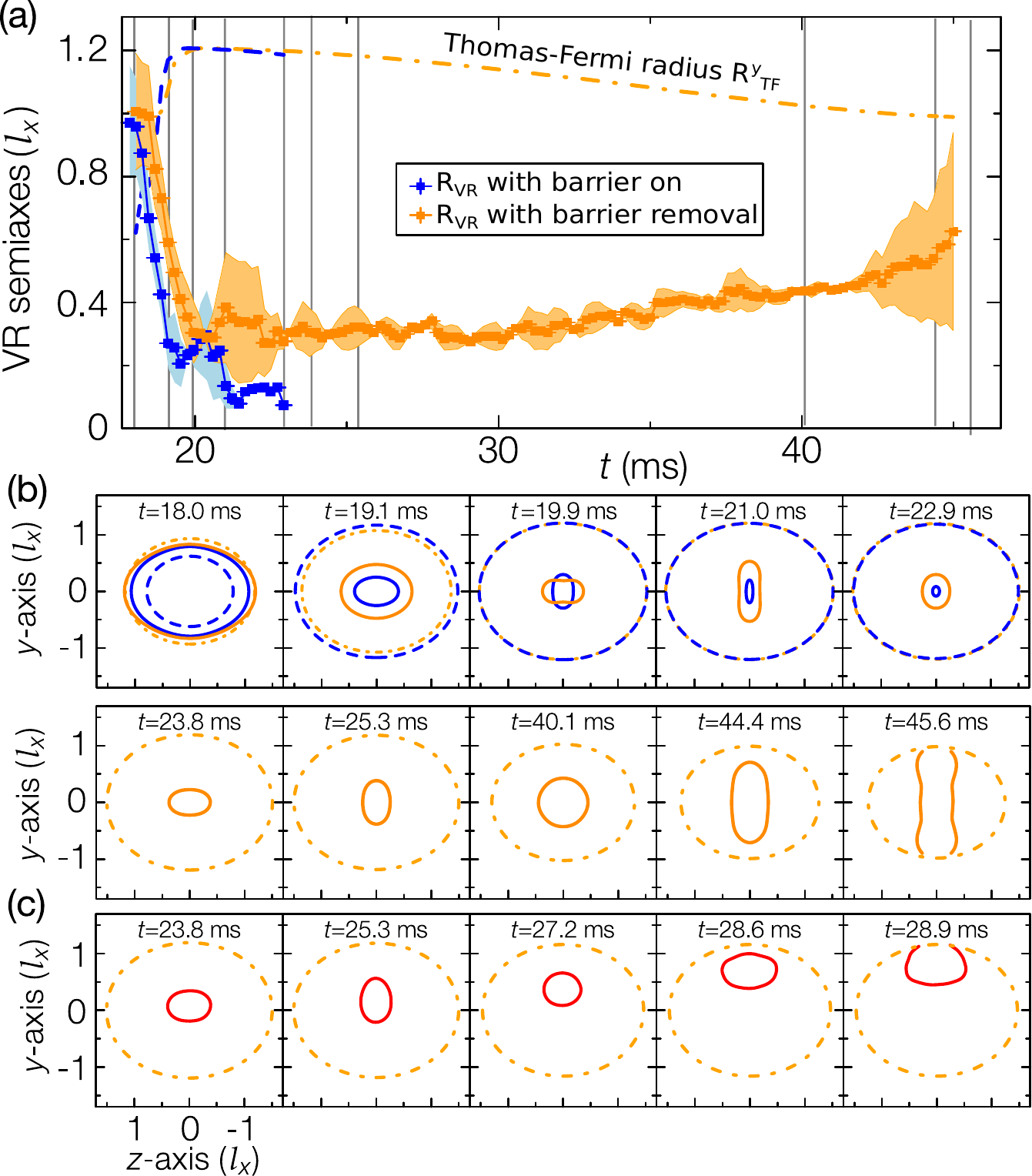}
\vspace{-0.4cm}
\caption{Vortex ring evolution under different conditions: (a) Evolution of the semi-axes and mean radius of the $4^{\rm th}$ VR ($z_0=0.25$, $V_0/\mu\simeq 0.8$), with barrier kept on (blue line) or removed during [13,53] ms (orange line). Shadowed areas mark limiting values of the two  semi-axes. Dashed blue/orange lines on top: transverse TF radius at the instantaneous VR location. (b) Dynamical 2D VR profiles with barrier on (blue) and removed (orange) plotted alongside the corresponding transverse TF surface (dash-dotted lines). Displayed profiles correspond to evolution times marked by vertical solid lines in (a), with the VR surviving only until $t \simeq 23$\,ms with barrier on.
(c) Typical evolution of a 2D VR profile in the case of barrier removal at $T\approx0.4\,T_c$. The VR moves off-axis, generating a single vortex handle at the boundary.}
\label{Fig4}
\end{center}
\end{figure}

This picture remains qualitatively correct for the probed $T \lesssim 0.4\,T_c$ (red symbols in Figs.~\ref{Fig2}-\ref{Fig3}), for which no thermal VR activation occurs ($k_BT < 0.8 V_0$). For a fixed BEC number,
superfluid flow, VR generation and early dynamics 
are not noticeably affected by the thermal cloud,
whose main effect is to add an extra potential to the BEC \cite{vortexT1, vortexT2, vortexT3,SM,ZNG2}. Over longer timescales, dissipation due to relative BEC-thermal motion becomes relevant, decreasing the VR lifetime (Fig.~\ref{Fig3}(b)).

To connect with Ref.~\cite{Liscience, Lidiss}, we implement in our simulations the same protocol by which vortices were observed in time of flight after gradually removing the barrier with a 40 ms linear ramp. The dynamics of the $4^{\rm th}$ VR generated in the same conditions as in Fig.~\ref{Fig2} [right VR in Fig.~\ref{Fig2}(e)] is shown in Fig.~\ref{Fig4}, including or excluding the barrier removal procedure. Upon removing the barrier (orange curve), the VR propagates for longer time and longer distance [Fig.~\ref{Fig4}(a)]. This facilitates the direct observation of Kelvin-wave oscillations [visible in Fig.~\ref{Fig4}(b)], whose period is consistent with the dispersion relation $\omega(k) \sim \kappa k^2/(4\pi) [\ln (2/(k\xi)-0.5772]$ \cite{barenghi-etal-1985, SM}. The longer lifetime can be attributed to the larger kinetic energy of VRs nucleated during the gradual barrier removal process. As the VR approaches the edge of the condensate, it breaks up into two anti-parallel vortex lines [Fig.~\ref{Fig4}(b), final snapshot] \cite{Piazza1, reichl-mueller-2013,wang-etal-2017}. Critically, thermal fluctuations destabilize the VR, 
causing it to drift off-axis, and reach the transversal boundary asymmetrically [Fig.~\ref{Fig4}(c)]. There, it reconnects with its image and forms a `vortex handle' \cite{Serafini2017,Levin, Mateo}.
This could explain why a single vortex line is typically detected in each experimental run after barrier removal \cite{Liscience,Lidiss}.

{\em Conclusions.}
We have studied the complex interplay between coherent and dissipative dynamics in a thin atomic Josephson junction. We have shown that resistive currents are directly connected with nucleations of vortex rings and their propagation into the superfluid bulk. In particular, dissipation originates from two irreversible effects: phonon emission when vortex rings are nucleated, and incompressible kinetic energy transfer from the superfluid flow to the swirling one of the nucleated vortex rings. 
The detailed understanding of the connection between vortex-ring dynamics and dissipation is valuable for advancing our comprehension of the complex superfluid dynamics in emerging atomtronic devices \cite{Amico_2017}.

{\em Acknowledgments.} We thank A. Smerzi and A. Mu\~{n}oz Mateo for valuable discussions. This work was supported by QuantERA project NAQUAS (EPSRC EP/R043434/1), EPSRC project EP/R005192/1, Fondazione Cassa di Risparmio di Firenze project QuSim2D 2016.0770, European Research Council grant agreement no.~307032 QuFerm2D and no.~637738 PoLiChroM, and European Union's Horizon 2020 research and innovation programme under the Marie Sk\l{}odowska-Curie grant agreement no.~705269.


%

\clearpage

\onecolumngrid
\begin{center}
\large Supplemental Material \\[2mm] \textbf{Critical transport and vortex dynamics in a thin atomic Josephson junction}\\[4mm]
\normalsize
{K. Xhani,$^{1,2}$ 
E. Neri,$^{3}$ 
L. Galantucci,$^{1}$ 
F. Scazza,$^{2,4}$ 
A. Burchianti,$^{2,4}$
K.-L. Lee,$^1$\\
C. F. Barenghi,$^{1}$
A. Trombettoni,$^{5}$ 
M. Inguscio,$^{2,4,6}$ 
M. Zaccanti,$^{2,3,4}$ 
G. Roati$^{2,4}$ and 
N. P. Proukakis$^{1}$}
\\[2mm]
\smallskip
{$^{1}$
Joint Quantum Centre (JQC) Durham-Newcastle, School of Mathematics, Statistics and Physics, \\Newcastle University, Newcastle upon Tyne NE1 7RU, United Kingdom\\
\mbox{$^{2}$European Laboratory for Non-Linear Spectroscopy (LENS), Universit\`{a} di Firenze, 50019 Sesto Fiorentino, Italy}\\
$^{3}$Dipartimento di Fisica e Astronomia, Universit\`{a} di Firenze, 50019 Sesto Fiorentino, Italy\\
$^{4}$Istituto Nazionale di Ottica del Consiglio Nazionale delle Ricerche (CNR-INO), 50019 Sesto Fiorentino, Italy\\
$^{5}$Istituto Officina dei Materiali del Consiglio Nazionale delle Ricerche (CNR-IOM) and Scuola Internazionale Superiore di Studi Avanzati (SISSA), Trieste, Italy\\
$^{6}$Department of Engineering, Campus Bio-Medico University of Rome, 00128 Rome, Italy}

\end{center}
\bigskip
\twocolumngrid
\setcounter{equation}{0}
\setcounter{figure}{0}
\setcounter{table}{0}
\setcounter{section}{0}
\setcounter{page}{1}
\makeatletter
\renewcommand{\theequation}{S.\arabic{equation}}
\renewcommand{\thefigure}{S\arabic{figure}}
\renewcommand{\thetable}{S\arabic{table}}
\renewcommand{\thesection}{S.\arabic{section}}

\section{Numerical Methods}
We study the superfluid transport of the molecular BEC through the thin barrier using two different models.
Specifically, the $T=0$ dynamics is modelled by the Gross-Pitaevskii equation. At finite temperatures we instead make use of a collisionless kinetic model, in which the condensate dynamics is self-consistently coupled to a dynamical thermal cloud described by a Boltzmann equation.

\subsection{Experiment Overview and System Parameters}
Superfluids of $N \simeq 10^5$ atom pairs of $^6$Li  are produced by cooling a balanced mixture of the two lowest spin states $|F=1/2, m_F=\pm 1/2 \rangle$ to $T/T_c\sim 0.3(1)$ \cite{Bur14,Liscience,Lidiss}.
Interactions between fermions are parametrized by $1/(k_F a)$, 
where $k_F = \sqrt{2 m E_F}/\hbar$ is the Fermi wave-vector ($m$ is $^6$Li atomic mass and $E_F$ the Fermi energy), and $a$ is the interatomic tunable $s$-wave scattering length. The focus of this work is on the regime of superfluidity of the molecular BEC, and we restrict our modelling to the case of $1/(k_Fa) \simeq 4.6$ .
To realize an atomic Josephson junction the fermionic superfluid is separated into two weakly-coupled reservoirs by focusing onto the atomic cloud a Gaussian-shaped repulsive sheet of light, yielding a trapping potential 
\begin{equation}\label{Vtrap}
\resizebox{.95\hsize}{!}{$V_\mathrm{trap}(x,y,z)=\frac{1}{2}M ({\omega_x}^2 x^2+ {\omega_y}^2 y^2+ {\omega_z}^2 z^2)+V_0 \cdot e^{\frac{-2x^2}{w^2}}$}
\end{equation}
where $\omega _{x,y,z}$ are the trapping frequency along $x$, $y$ and $z$-directions, $M=2m$ is the molecular mass, $V_0$ is the height of the Gaussian barrier and $w \approx  2.0 \pm 0.2\,\mu$m is the barrier $1/e^2$ width, which is just four times wider than the superfluid coherence length $\xi$.
The experimental trap frequencies are $\omega_x\simeq2 \pi \times 15$ Hz,  $\omega_y \simeq2 \pi \times 187$Hz, $\omega_z\simeq 2 \pi \times 148$Hz (cigar-shaped trap), and $V_0$ is varied in the regime of $0.6\, \mu \lesssim V_0 \lesssim 1.2\, \mu $  where $\mu$ denotes the chemical potential of the system.

%

\subsection{Gross-Pitaevskii Equation ($T=0$)}

At $T=0$ we model the system by the molecular BEC wavefunction $\psi$ obeying the time-dependent Gross-Pitaevski equation (GPE):
\begin{equation}\label{GPE}
i \hbar \frac{\partial \psi (\vec{r},t) }{\partial t}=\left(-\frac{\hbar ^2}{2M} \nabla ^2 + V_\mathrm{trap}  + g |\psi(\vec{r},t)|^2 \right) \psi(\vec{r},t)  
\end{equation}
where $g=4\pi \hbar ^2 a_{M}/M$ is the interaction strength, and $a_\mathrm{M}=0.6\, a\simeq 7 \times 10^{-3} \, l_\mathrm{x}$ is the molecular scattering length. The equilibrium state is found by substituting $\psi(\vec{r},t)=\psi _0(\vec{r}) \exp{(-i \mu t/ \hbar)}$ which gives the time-independent GPE:
\begin{equation}
\mu \psi(\vec{r})=\left( -\frac{\hbar ^2}{2M} \nabla ^2  + V_\mathrm{trap} + g |\psi(\vec{r})|^2 \right)  \psi_0(\vec{r})
\end{equation}
with $\mu$ the system chemical potential.
The equilibrium state is obtained numerically via imaginary time propagation in the presence of an {\em additional} linear potential, $-\epsilon x$, along the $x$-direction which sets up the desired initial population imbalance, $z_0$, between the two wells (i.e.~initial chemical potential difference). As there is initially a larger population in the right well, the initial flow is induced along the negative $x$-direction. The BEC dynamics instead is initiated by the instantaneous linear potential removal at $t=0$. Eq.~\ref{GPE} is studied in dimensionless form, with length scaled to the harmonic oscillator length along the $x$-direction, $l_\mathrm{x} =\sqrt{\hbar/m \omega_\mathrm{x}} \simeq 7.5 \, \mu $m.
In our numerical simulations we use grid sizes $\left[-24,24\right]l_x,\left[-4,4\right]l_x, \left[-4,4\right]l_x$ along the $x$, $y$ and $z$-directions, and  $1024 \times 128 \times 128$ grid points respectively. Throughout this work, the barrier width is set to $w\simeq2\mu$m $\approx 4\xi$, with $\xi = 1 / \sqrt{8 \pi a_M n_\mathrm{max}}\approx 0.5 \, \mu$m$\, \simeq0.067 \, l_\mathrm{x}$.   
To account for experimental numbers, Fig.~1(a) of the main paper considered molecule numbers in the range $(6-12) \times 10^4$. Throughout this Supplemental Material instead, we fix the condensate number at $N_\mathrm{BEC}=6\times10^4$. 
In analyzing our results, we express the barrier height $V_0$ in units of the system chemical potential $\mu$. The numerically-extracted equilibrium $\mu$ is well approximated by the analytical formula in the Thomas-Fermi approximation:
\begin{equation}
\mu= \frac{1}{2} \hbar \overline{\omega} \left( \frac{15 N_\mathrm{BEC} a_M}{\overline{l}}\right) ^{2/5}
\label{mu}
\end{equation}
with $\overline{\omega}=(\omega _x \omega _y \omega _z)^{1/3}$ and $\overline{l}=\sqrt{\hbar/M \overline{\omega}}$ the geometric mean of harmonic oscillator lengths \cite{ketterle}.
For typical parameters ($T=0$, $N_\mathrm{BEC}=6 \times 10^4$ molecules) $\mu \simeq 114 \hbar \omega _x$ and the barrier height numerically explored is in the range $[0.6,1.22] \, \mu$. 

\subsection{The Two-mode model}

In this section we discuss the validity of the commonly-used two-mode model for the specific geometry and parameter regime of our system.

The two-mode model cannot capture the observed dynamical regimes, since -- with increasing population imbalance -- it always predicts a transition  from the Josephson oscillation regime to the Macroscopic Quantum Self-Trapping (MQST) regime, in which the population imbalance oscillates around a non-zero value.

However, as already clearly demonstrated in the main paper, for the experimental parameters probed here -- namely a narrow barrier of width $w/\xi=4$ and a relatively low barrier height $V_0/\mu$ in the range [0.6, 1.22] -- we have instead numerically found a transition from the Josephson regime (for small initial population imbalance) to a distinct, dissipative, regime associated with defect generation. Typical dynamics of the population imbalance in these two regimes are shown in the two insets to Fig.~1(a) of the main paper. Such numerical results are confirmed by their excellent agreement with the experimental data over the entire range of barrier height explored (see Fig.~1(a)). 


Even though the two-mode model fails to capture the nature of the dynamical regime to which the system transitions with increasing initial population imbalance, it is nonetheless interesting to examine the extent to which such a model can predict {\em either} the location of the transition region where Josephson plasma oscillations no longer occur, {\em or} the actual frequency of the Josephson oscillations within the Josephson regime.

To address such issues, we start with a brief overview of the two-mode model (\cite{MQST1, MQST3}).


Assuming the two superfluids are well-localized in each well, the system wavefunction can be written as a linear superposition of the left and right condensate wavefunctions as
\begin{equation}
\psi(\vec{r},t) = \psi_{L} (t) \cdot \eta_{L} (\vec{r})+\psi_{R} (t) \cdot \eta_{R} (\vec{r})
\end{equation}
where $\psi_{L}(t)=\sqrt{N_{L}} e^{i \phi_{L}}$ and  $\psi_R(t)=\sqrt{N_{R}} e^{i \phi_{R}}$. Here
\begin{equation}
\int \eta_i \cdot \eta_j d\vec{r}=\delta_{i,j} 
\end{equation}
with $i$, $j$ denote left, right, and $N_{L(R)}$ or $\phi _{L(R)}$ are the left (right) condensate number and phase. 
This simple model is called the `two-mode model' as in this approximation only the ground state and the first excited state are populated. If  we define the population imbalance  
\begin{equation}
z_{BEC}=\frac{N_{L}-N_{R}}{N}
\end{equation}
(where $N=N_{L}+N_{R}$),
 and the phase difference $\Delta \phi =\phi_{L}-\phi_{R}$, the system energy in the two-mode model takes the form

\begin{equation}
E= \frac{UN}{2} z^2-2K \sqrt{1-z^2} \cos \left( \Delta \phi \right).
\end{equation}
where $K$ is the tunneling energy and $U$ the onsite interaction energy.
For $z_0 \ll 1$, this can be written as
\begin{equation}
E= \frac{UN}{2} z^2-2K \cos \left( \Delta \phi \right) \;.
\end{equation}

The condition for obtaining MQST corresponds to having an initial energy $E_0=E(z_0,\Delta \phi = 0) > E_{cr}$, where $E_{cr}=E(z=0, \Delta \phi = \pi)$. Thus the critical imbalance for obtaining MQST is:
\begin{equation}
z_{cr}=\sqrt{\frac{8K}{UN}} \;.
\end{equation} 
\begin{figure}[t!]
\centering
  \includegraphics[width=.8\linewidth]{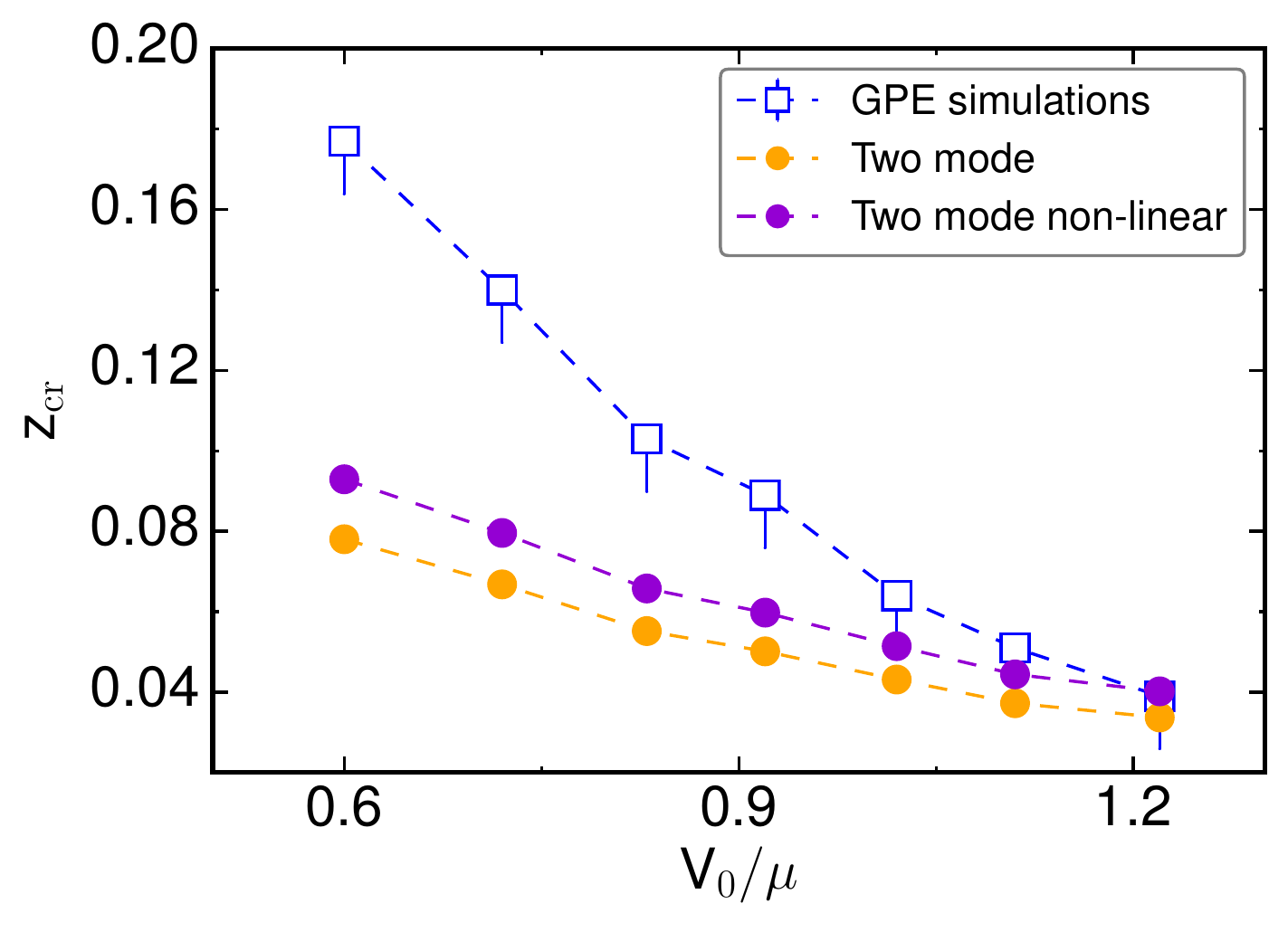}
  \caption{Critical initial population imbalance for different barrier height $V_0/\mu$ for 6x10$^4$ particles extracted from GPE simulations (blue squares -- also shown in Fig. 1(a) of main paper), from linear (orange circles) and nonlinear (violet circles) two-mode model.
Note that while all models agree on the {\em existence} of a Josephson regime below their corresponding predictions for $z_{cr}$, the two-mode models predict a transition to a different regime (namely MQST) than that found in the GPE simulations (dissipative regime).
}
  \label{fig:zc}
\end{figure}
This condition corresponds to populating the first excited state of the system, which is the antisymmetric state. Thus the tunneling energy $K$ can be extracted from the difference  between the antisymmetic-state energy,  $E_a$, and the symmetric-state energy, $E_s$, via
\begin{equation}
2K=\frac{\left(E_a-E_s \right)}{N} \;.
\end{equation}
Hereafter we use $E_{J}$, defined as $E_{J}=2K$.
The onsite interaction energy $U$ can be calculated {\em either} from\\
\noindent the linear two-mode model \citep{MQST1}, via 
\begin{equation}
U=g \int \eta_{L}^4 d\vec{r}
\end{equation}
{\em or}
from the nonlinear two-mode model \cite{andrea2003} as:
\begin{equation}
U_{NL}=2 \frac{\partial \mu}{\partial N}
\end{equation}

To evaluate the two-mode model prediction for the location of the crossover between Josephson and MQST, we calculate $z_{cr}$ both from the linear and from the nonlinear two-mode model for each $V_0/\mu$, and plot it against $V_0/\mu$ in the experimentally relevant range $[0.6,1.22]$ in Fig.~ \ref{fig:zc}.
We also plot on this figure the corresponding extracted values of $z_{cr}$ from the GPE simulations [noting that in the latter case, the transition is to a different, dissipative, regime, rathen than to MQST].
We find that both linear and nonlinear two-mode model clearly significantly underestimate the numerically predicted critical value of $z_{cr}$ for all $V_0 / \mu \le 1.1$, yielding a correct value for the transition only at the highest probed point $V_0/\mu=1.22$ (see Fig.~\ref{fig:zc}). 

Next, we examine the extent to which the oscillation frequency of the Josephson oscillation is correctly predicted by such two-mode model.\\
%
 %

 \begin{figure}[!h]
  \centering
  \includegraphics[width=.8\linewidth]{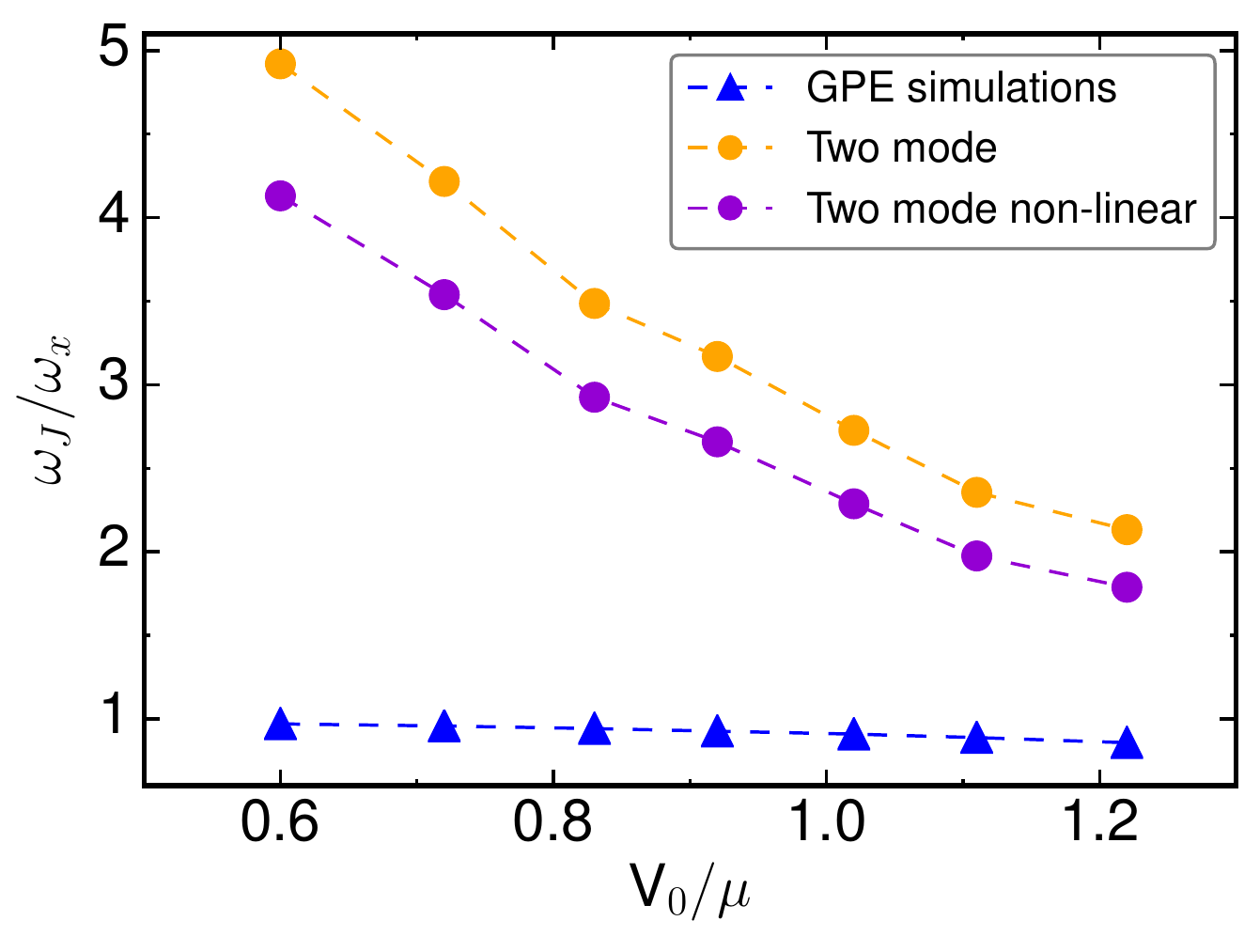}
  \caption{The oscillation frequency of the population imbalance for $z_0< z_{cr}$, i.e. Josephson plasma frequency, from linear (orange circles) and nonlinear (violet circles) two-mode model and GPE simulations (blue triangles).}
\label{fig:nuJ}
\end{figure}
In the regime of the Josephson plasma oscillation, where $z_0<z_{cr}$, and for $E_J \ll E_c$ the two-mode model prediction for the oscillation frequency is:
\begin{equation}
\omega_J=\frac{\sqrt{E_J E_c}}{\hbar}=\frac{\sqrt{E_J U N}}{\hbar} \;.
\end{equation}
Plotting $\omega_J$ against $V_0/\mu$ in the considered range (Fig.~\ref{fig:nuJ}), we find the predictions of both linear and nonlinear two-mode model to be incorrect by at least a factor of 2 (for $V_0/\mu = 1.22$), with the disagreement rapidly increasing for lower values of $V_0/\mu$.

The above analysis thus proves that -- for the parameter regimess of the experiment under study -- the two-mode model is out of its validity range, due to the considered values of the ratio
$V_0/\mu$ and the narrow width of the junction.

\subsection{The Collisionless “ZNG” Kinetic Model ($T>0$)}
At finite temperature, the bosonic quantum gas is partially condensed and we must consider the presence of the thermal cloud. 
The GPE is therefore generalized to account for the thermal cloud mean field potential, $2g n_\mathrm{th}$, so that Eq.~\ref{GPE} becomes \cite{ZNG2}:
\begin{equation}
\label{gped}
i \hbar \frac{\partial \psi}{\partial t}=\left[ - \frac{\hbar^2 \nabla ^2}{2M}+V_\mathrm{trap}+g(|\psi|^2+2n_\mathrm{th})\right] \psi \;.
\end{equation}
The equilibrium $\psi _0$ solves the time-independent generalized GPE:
\begin{equation}\label{gped2}
\mu(T) \psi _0=\left( -\frac{\hbar ^2}{2M} \nabla ^2  + V_\mathrm{trap} + g (|\psi _0 |^2 + 2 n^0 _\mathrm{th})\right)  \psi_0
\end{equation}
where $n^0 _\mathrm{th}$ is the equilibrium thermal cloud density, while $\mu(T)$ is the temperature-dependent system chemical potential counting for the thermal cloud equilibrium mean field potential. This has been used for extracting $V_0/\mu(T)$ ($x$-axis of phase diagrams in Fig. 1 of the main paper). 
To account for thermal cloud dynamics, we solve this equation self-consistently with a collisionles Boltzmann equation for the thermal molecule phase-space distribution, $f$, obeying: 
\begin{equation}
\label{bolt}
\frac{\partial f}{\partial t}+ \frac{\vec{p}}{M} \cdot \nabla_{\vec{r}} f-\nabla_{\vec{r}}V_{eff} \cdot \nabla_{\vec{p}}f=0
\end{equation}
where $V_{eff}=V_\mathrm{trap}+2g [|\psi|^2+n_{th}]$ is the generalized mean-field potential felt by the thermal molecules, and the thermal cloud density is defined by 
\begin{equation}
n_{th}=\frac{1}{(2\pi\hbar)^3} \int d\vec{p} ~ f(\vec{p},\vec{r},t) \;.
\end{equation}
The initial finite-temperature equilibrium distribution is obtained iteratively for a fixed total atom number, as described in Ref.~\cite{ZNG2, ZNG3}.
Our model corresponds to the collisionless limit of the ``Zaremba-Nikuni-Griffin'' (ZNG) kinetic theory which has been successfully used to model collective modes
, vortex dynamics and evaporative cooling \cite{vortexT1, ZNG2, nick_book, vortexT2, vortexT3}.

In choosing parameters for our finite temperature simulations, we ensure that the BEC number is equal to the corresponding $T=0$ number, fixed here to $6 \times 10^{4}$ particles.
In order to capture the entire thermal cloud -- which resides primarily outside the BEC region -- our finite temperature simulations use an extended grid length  $\left[-48,48\right]l_x,\left[-8,8\right]l_x, \left[-8,8\right]l_x$ along the $x$, $y$ and $z$ directions respectively, and $2048 \times 256 \times 256$ grid points for the thermal cloud.
\section{Flow Dynamics: Superfluid Current, Phase Slips, Backflow and Vortex Ring Energy}
\subsection{Extracting the superfluid current through the barrier}
\begin{figure}[h!]
\begin{center}
\includegraphics[width=0.99 \columnwidth]{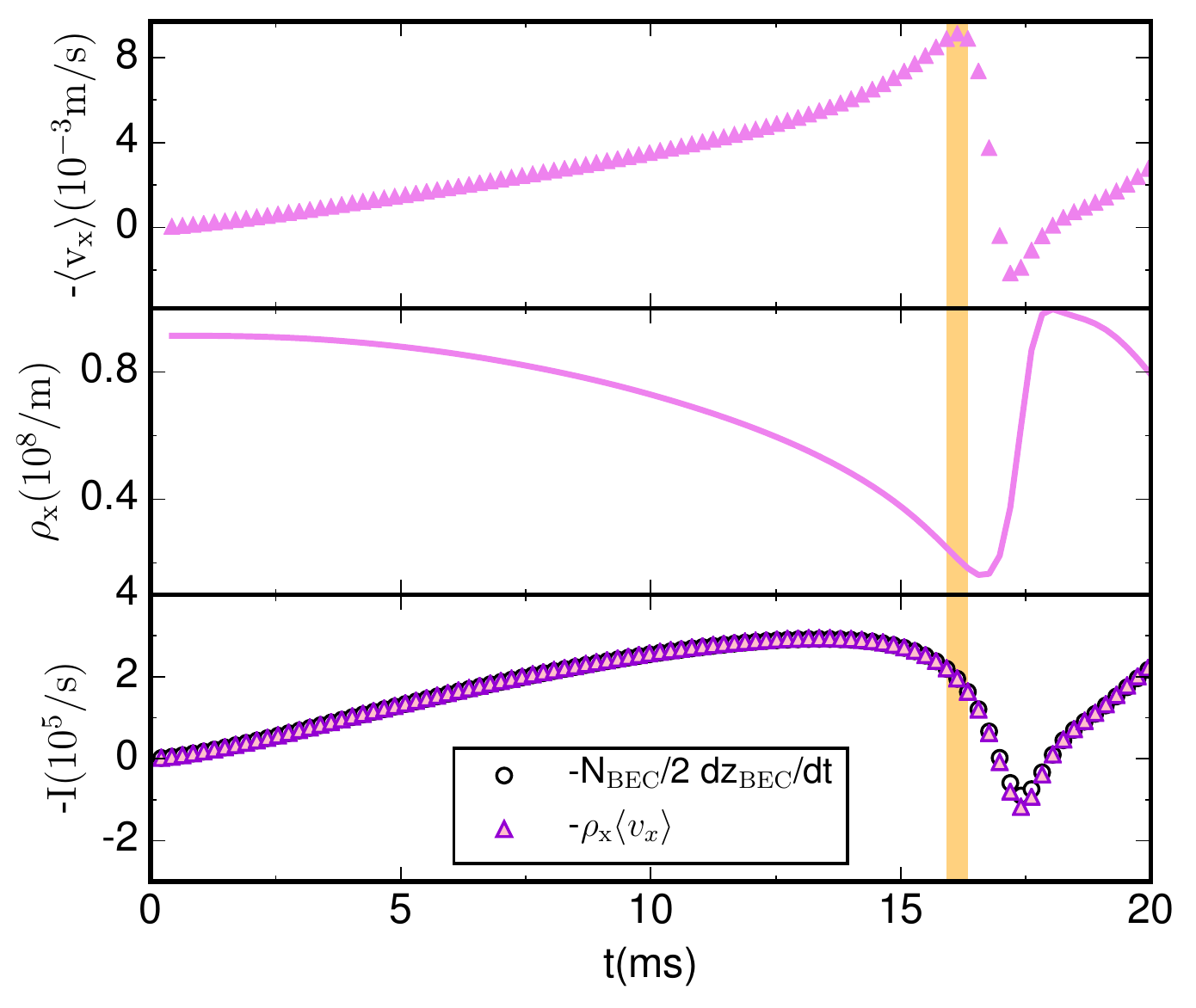}
\caption{Density-weighted superfluid velocity along the x-direction (upper plot), the transverse integrated density (middle plot), and their product (violet symbol lower plot) as a function of time calculated at the trap center for $V_0/\mu\simeq 0.8$ and $z_0=0.13$. In the lower plot we also show the current calculated from the time derivative of the population imbalance (black symbols). Vertical shaded area highlight the maximum value of $-\left\langle v_x \right\rangle$. } 
\label{vx}
\end{center}
\end{figure}

Figs.~1 of the main paper show numerical results for the maximum superfluid current and its temporal profile. Here we show how these results have been obtained.
There are two ways to calculate the superfluid current: the first one is from the time derivative of the population imbalance $I=\dot{z}_{\rm BEC} N_{BEC}/2$ and the second one from the transverse integral of the probability current density 
\begin{equation}
I= \int_{R^z _{TF}}\int_{R^y _{TF}} j_x(x=0,y,z) \, dy \, dz\;,
\end{equation}
where $R^{y(z)} _\mathrm{TF}$ is the Thomas-Fermi radius along the $y(z)$-direction. 
Here $j_x$ is the $x$-component of the density current of probability defined as:
\begin{equation}
\bold{j}=\frac{\hbar}{2iM} (\psi ^* \nabla \psi- \psi \nabla \psi ^*)
\end{equation}
The latter can also be written as:
\begin{equation}
I=\int_{R^z _\mathrm{TF}}\int_{R^y _\mathrm{TF}} |\psi(0,y,z)|^2 \cdot v_x(x=0,y,z) dy dz
\end{equation}
where $v_x$ is the component of the superfluid velocity along the $x$-direction.
By defining the density-weighted superfluid velocity as:
\begin{equation}
\left\langle v_x \right\rangle=\frac{\int v_x \cdot |\psi(x=0,y,z)|^2  dy dz}{ \int |\psi(0,y,z)|^2 dy dz} \;, 
\end{equation} 
we can write 
\begin{equation}
I=\rho _x \left\langle v_x \right\rangle \label{I-rho} \;,
\end{equation}
where $\rho _x= \int |\psi|^2(x=0,y,z)dy dz$.
The numerical reconstruction of Eq.~(\ref{I-rho}) is shown in Fig.~\ref{vx}.
These two ways of calculating $I$ are equivalent as shown in Fig.~\ref{vx} (lower plot) for the case of $V_0/\mu\simeq 0.8$ and $z_0=0.13$. We note here that the maximum of $-\left\langle v_x \right\rangle$ is shifted with respect to the current maximum due to the varying density (Fig.~\ref{vx}).

\begin{figure}[h!]
\begin{center}
\includegraphics[width=0.8 \columnwidth]{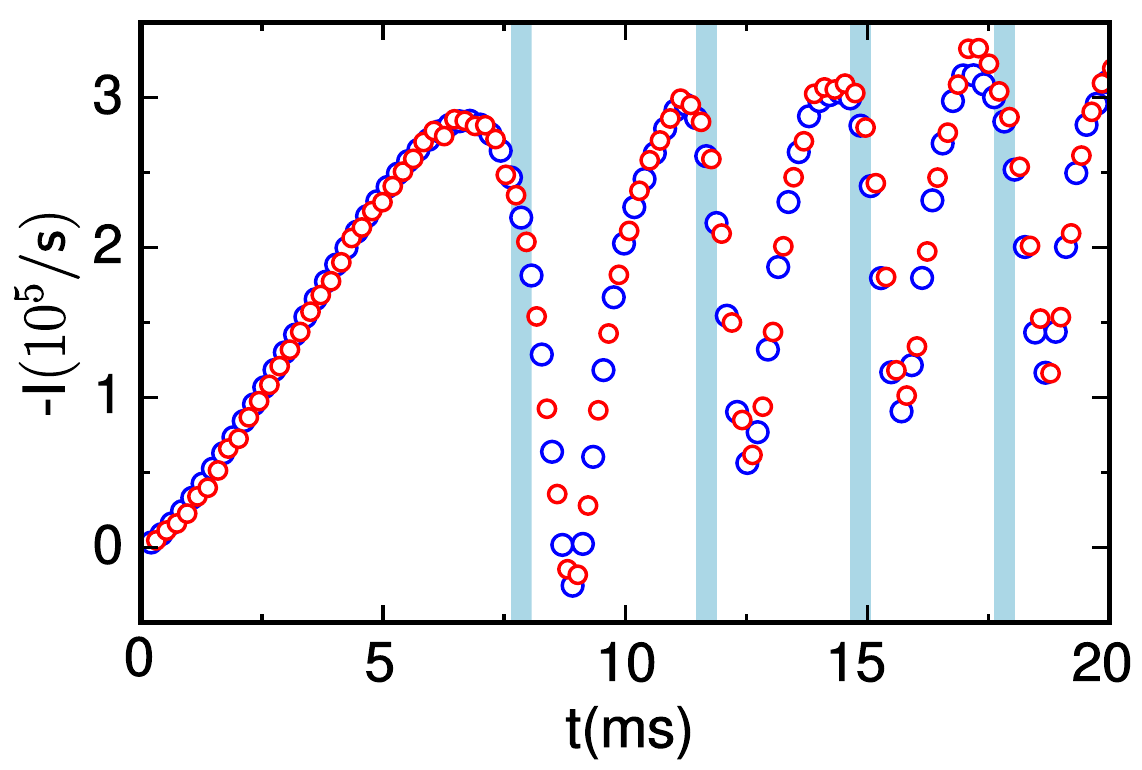}
\caption{The temporal evolution of the superfluid current calculated from the time derivative of the population imbalance for $V_0/\mu\simeq 0.8$ and $z_0=0.25$. Shown are the T=0 (blue symbols) and T=0.4 T$_\mathrm{c}$ results (red symbols). Vertical shaded areas highlight the maximum values of $-\left\langle v_x \right\rangle$ already shown in Fig.~2 of the main paper.} 
\label{I_z025}
\end{center}
\end{figure}

To further support the statement made in the main text that the (experimentally relevant) small thermal cloud fraction does not noticeably influence the VR nucleation process and early dynamics, we also show in Fig.~\ref{I_z025} the superfluid current calculated from the time-derivative of the population imbalance (as in Fig.~\ref{vx}(c)), but for a larger population imbalance $z_0 = 0.25$, corresponding precisely to the parameter regime of Fig.~2 of the main paper.
The superfluid current is seen to be unaffected by temperature, even in the case when multiple rings are generated.


\subsection{Second order term in the Josephson current-phase relation}
Here we explain the role of the second order harmonic to the current-phase relation plotted by the green shaded area in Fig.~1(b) of the main text. 

The overall trend of the maximum current as a function of $V_0/\mu$ is quantitatively captured by extending the analytic model developed in Ref. \cite{Ic2} for homogeneous Bose gases and rectangular barriers, to the harmonically trapped case with a Gaussian barrier, relevant for our study.
While more details will be given elsewhere, in the following we briefly summarize how such model extension has been obtained.
First, we recall that based on a perturbative approach valid in the limit $V_0 \gg \mu$, Meier and Zwerger \cite{Ic2} derived analytic expressions for the critical current, up to second order in the tunneling hamiltonian, for a homogeneous, $T=0$ weakly interacting BEC. 
In particular, they obtained explicit predictions for the first and second order contributions to the superfluid current, respectively denoted by $I_c$ and $J_1$, yielding a current-phase relation of the kind \cite{Ic2}:
\begin{equation}
I(\varphi)=I_c {\rm sin}(\varphi)+ J_1 {\rm sin}(2\varphi) \equiv I_c \cdot ({\rm sin}(\varphi)+\bar{g} \cdot {\rm sin}(2\varphi))
\label{CPRel}
\end{equation}
where $\bar{g}=J_1/I_c$.
Notably, these terms solely depend upon the bulk condensate density and the boson tunneling amplitude, which in turn can be recast in terms of the bulk chemical potential and the single-particle transmission coefficient across the barrier \cite{Ic2}.
As such, within this framework, predictions for the maximum current supported by a generic junction can be obtained from the knowledge of the bulk properties of the superfluid, and by evaluating the single-particle transmission coefficient associated with the specific barrier under consideration.
\begin{figure}[h!]
\begin{center}
\includegraphics[width=0.8 \columnwidth]{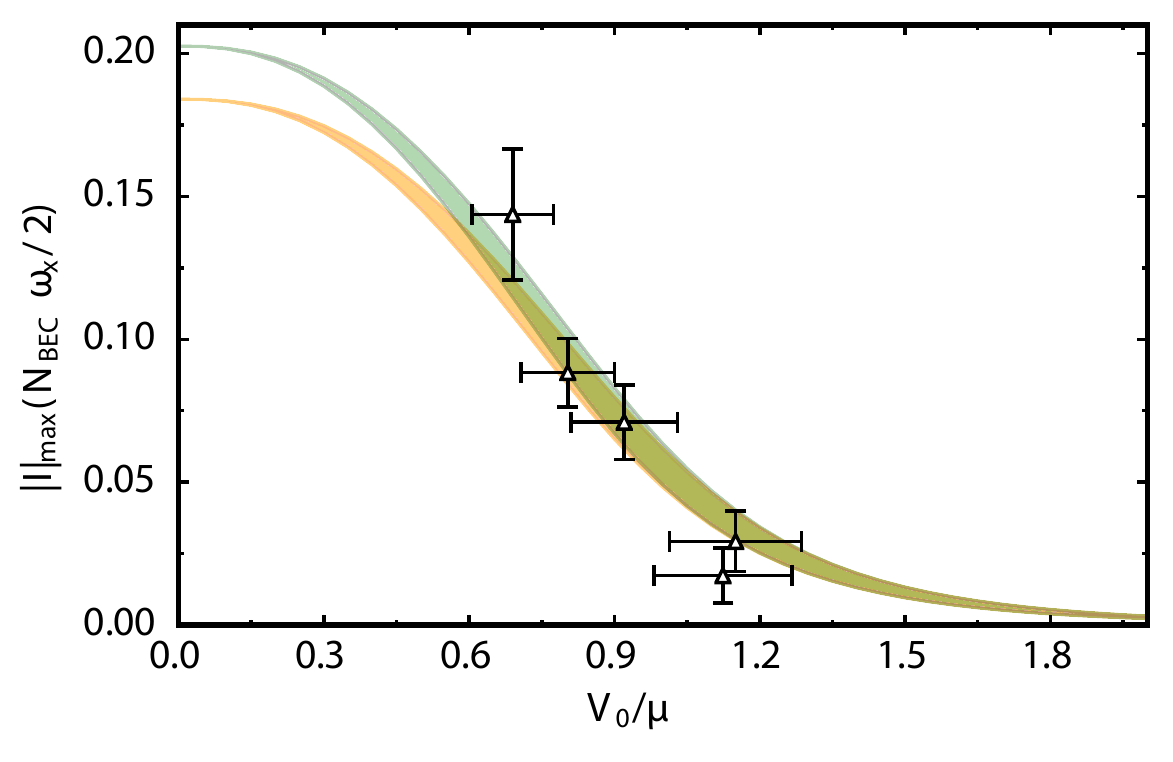}
\caption{The prediction of the maximum superfluid current taking only the first order term in the current-phase relations (yellow profile) and including also the second order terms (green profile) and the experimental data (black points).} 
\label{I}
\end{center}
\end{figure}

To proceed with obtaining the theoretical curves, we modified the analytic results given in Ref. \cite{Ic2} in the following ways.
First, we derived an analytic expression for the tunneling amplitude $t_{cc}(V_0, \mu)$ (using the notation of \cite{Ic2}) across our Gaussian barrier, by approximating the Gaussian profile with a symmetric Eckart potential of the kind $V_{Ec}=V_0/{\rm cosh}(x/d)^2$ with $d=0.6w$.\\
Second, we employed local density approximation to recast the first and second order currents within an integral form, in order to account for the inhomogeneous density distribution featured by our trapped samples.
From the knowledge of $I_c$ and $J_1$, we then obtained the maximum current enabled by our junction, up to first and second order, respectively. These are shown in Fig.~\ref{I}. In the former case, the critical current is simply given by $I_c$.

The yellow shaded area in Fig. \ref{I} is delimited by the  trend of $I_c$ predicted by our analytic model, assuming a $\pm 5 \%$ relative variation of the peak chemical potential $\mu_0$ of the trapped Bose gas at $T=0$, around the nominal value of $\mu_0$ based on the measured molecule number and trapping frequencies.  

While $I_c$ provides an excellent description of $I_\mathrm{max}$ derived from the experimental data and numerical simulations for $V_0/\mu \geq 0.9$, it systematically underestimates the maximum current found for lower barrier heights.  
This mismatch is expected in light of the fact that second order contributions become increasingly important for progressively lower $V_0/\mu$ values \cite{Ic2}.
In order to account for second order corrections to $I_\mathrm{max}$, we exploited our model prediction for $J_1$ in connection with the analytic results obtained in Ref. \cite{Goldobin2007} for a generic current-phase relation with first and second harmonics (Eq. (\ref{CPRel})). 
For any value of $\bar{g}=J_1/I_c$, it can be shown \cite{Goldobin2007} that the maximum current will read:
\begin{equation}
\frac{I_\mathrm{max}}{I_c} = \frac{(\sqrt{1 + 32 \bar{g}^2} + 3)^{3/2} (\sqrt{1 + 32 \bar{g}^2} -1)^{1/2}}{32 |\bar{g}|}
\label{IMax}
\end{equation}

The green shaded area in Fig.~1(b) of the main paper shows the trend of $I_\mathrm{max}$ based on Eq.~(\ref{IMax}) and the value of $\bar{g}$ derived from our analytic model, assuming  a $\pm 5 \%$ variation of the peak chemical potential, as for the first order case.
By comparing corresponding first (yellow area) and second order (green area) contributions in Fig.~\ref{I}, one can notice how inclusion of second harmonics generally increases $I_\mathrm{max}$, and enables to excellently reproduce both experimental and GPE results, down to barrier heights as low as $V_0\sim 0.6 \mu$.
 
\subsection{Population imbalance decay and phase slippage}

Here we provide more evidence for interpreting the relation between population imbalance decay (dissipation), vortex ring (VR) nucleation \cite{VRnucl1, VRnucl3, VRnucl2}, and phase slippage. 

\subsubsection{VR effects on the population imbalance profile}

\begin{figure}[h!]
\begin{center}
\includegraphics[width=.7 \columnwidth]{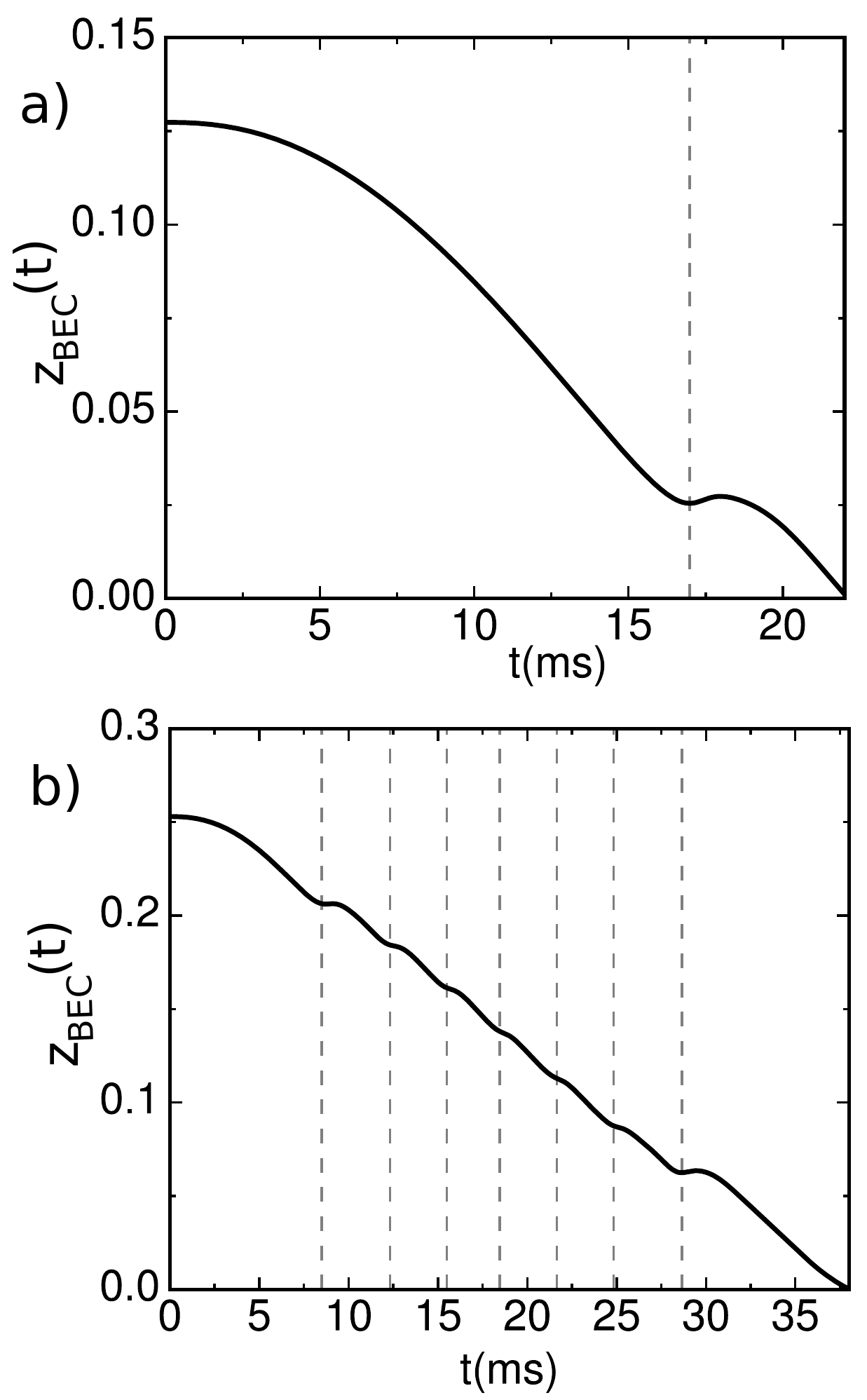}
\caption{Temporal evolution of the population imbalance for $z_0=0.13$ (a) and $z_0=0.25$ (b) both for $V_0/\mu \simeq 0.8$. The dashed grey line show the moment at which different VRs enter the Thomas-Fermi surface corresponding to the grey-areas shown in the Fig. 2 of the main paper.
} 
\label{zbec}
\end{center}
\end{figure}

Fig.~1(a) of the main text plots the phase diagram in terms of a critical population imbalance $z_{\rm cr}$. This value was identified as the value of $z_0$ at which $z_{BEC}(t)$ firstly exhibits decay of the population imbalance to zero. 
Fig. \ref{zbec} shows the evolution of $z_{BEC}(t)$ for $V_0/\mu\simeq 0.8$ and population imbalances (a) $z_0=0.13$ and (b) $z_0=0.25$. 

As discussed in the main text, such dips in the otherwise rapid (and monotonic) decay of $z_{BEC}(t)$ is due to the generation of VRs, which are nucleated outside of the Thomas-Fermi surface and subsequently enter such surface -- with the corresponding times indicated by vertical dashed line in Fig. \ref{zbec}.
Specifically for $z_0=z_{cr}$ there is only one VR generated. 
For $z_0=0.25$, the $z_{BEC}(t)$ achieves its zero value later in time with respect to $z_0=0.13$. This is not just because its value is higher but also because a VR causes a backflow (see next section) everytime it is generated, i.e.\ larger number of VRs give a cumulative effect slowing down the population imbalance decay.  
Fig.~\ref{zbec}(b) shows that a higher population imbalance leads to a larger number of vortex rings generated (an effect already shown in Fig.~3(b) of the main paper). Specifically, seven VRs are generated in the case $z_0=0.25$ (case discussed in Fig.~2 of main paper). The total number of VRs ($N_\mathrm{VR}$) propagating in the left well, which is plotted in the main paper Fig. 3(b) is found by looking at the 3D density plots. However different methods can be used as comparison to check for consistency: $N_\mathrm{VR}$ can be also found by counting the number of -$\left\langle v_x\right\rangle $ time evolution maxima. Another way to stimate $N_\mathrm{VR}$ is by the expression \cite{Anderson}:
\begin{equation}
\langle \mu _L -\mu _R \rangle= h \langle \frac{dn}{dt}  \rangle
\label{NVR}
\end{equation}
where the $\langle \cdots \rangle$ indicate time-averaged values, $(\mu_ L- \mu _R)$ is the chemical potential difference and $dn/dt$ is the rate of VRs crossing a certain path. The chemical potential in the left (right) well is estimated by using Eq.~(\ref{mu}), upon replacing N$_{BEC}$ with $N_L= (1-z_\mathrm{BEC})/2$, ($N_R= (1+z_\mathrm{BEC})/2$). 
\begin{figure}[h!]
\begin{center}
\includegraphics[width= \columnwidth]{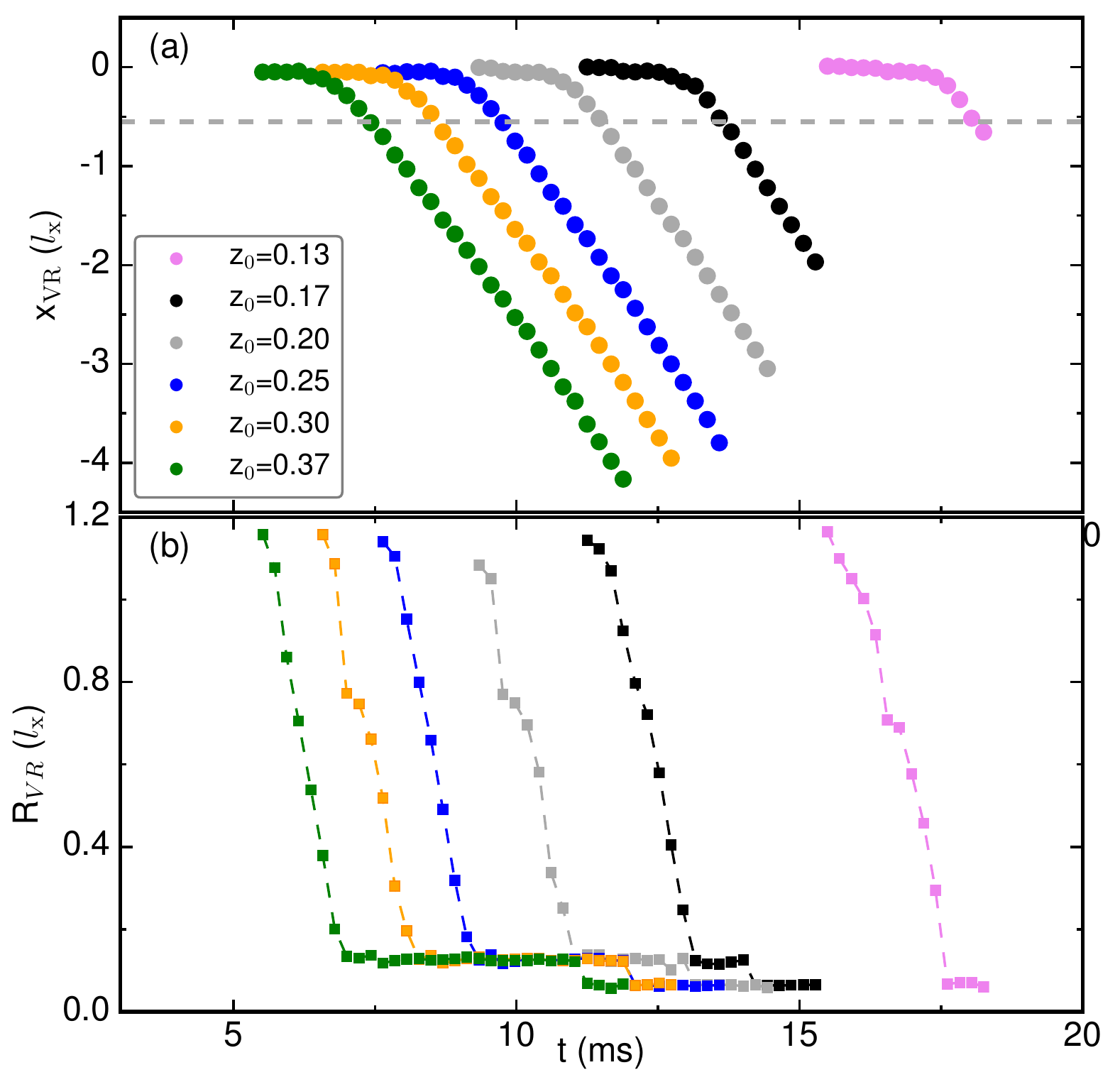}
\caption{Time evolution of (a) the first vortex ring axial position, and (b) of its semiaxes mean value, generated for different initial imbalances. Horizontal grey line in (a) at $x = - 0.55 l_x$ indicates the temporal value from which we start extracting the linear fit, $v_{VR}$ (shown in Fig.~3 of main paper.)} 
\label{1stVR}
\end{center}
\end{figure}

\begin{figure*}[tb] 
\centering
 \makebox[\textwidth]{\includegraphics[width=.5\paperwidth]{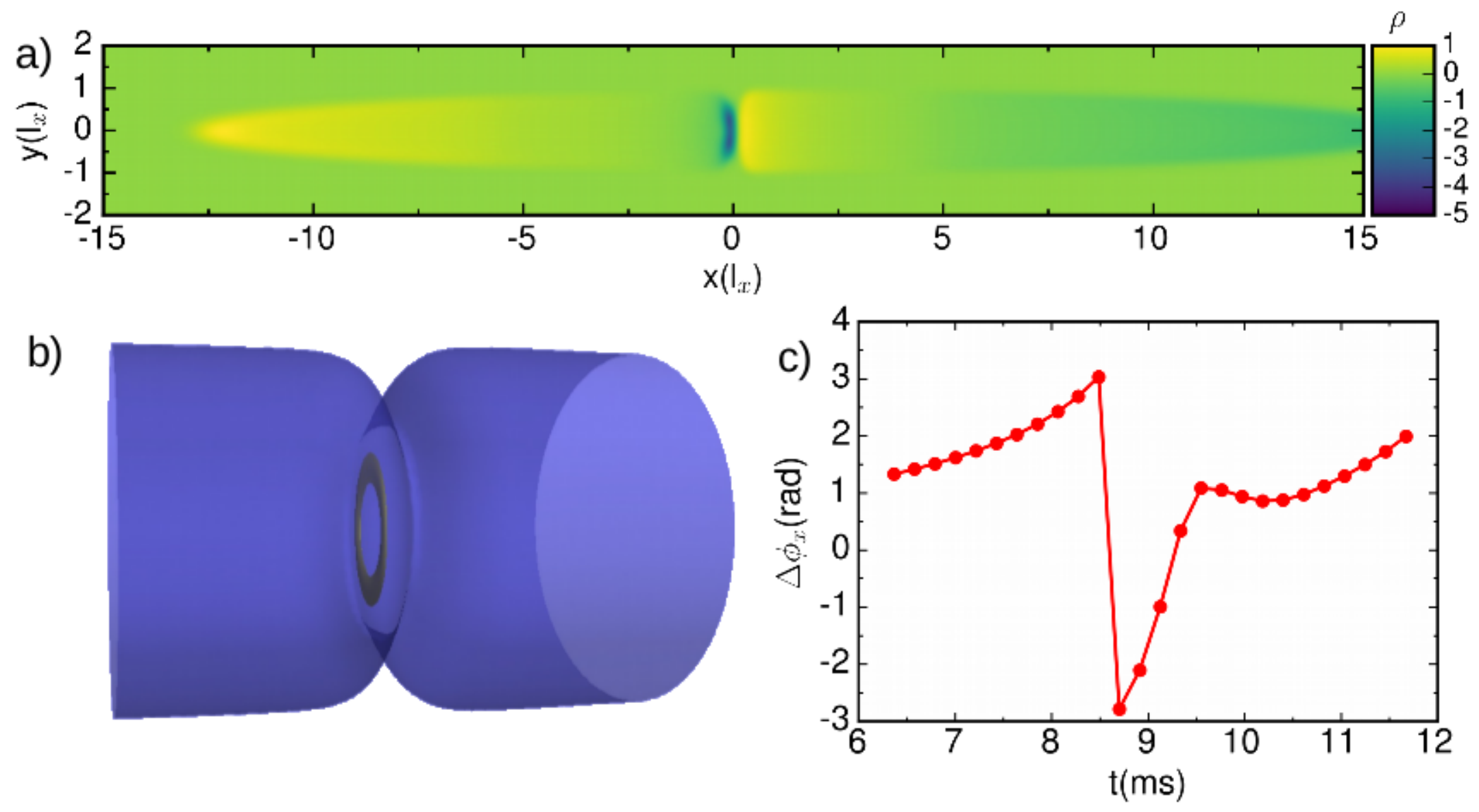}}
\caption{Relation between phase slip and VR generation:
(a) 2D BEC density after substracting the background 
density, scaled to its maximum value, and
(b) 3D density profile (density isosurface taken at 5\% of maximum density) showing the VR at t=8.7 ms.
(c) Corresponding relative phase evolution in time.
Example here is for
$V_0/\mu\simeq 0.8$ and $z_0=0.25$ 
(parameters of Fig.~2 of main paper).
}
\label{slippage1}
\end{figure*}

Note that the first VR is generated earlier in time with higher population imbalance, as shown in Fig.~\ref{1stVR}: 
This is because the larger $z_0$, i.e. larger chemical potential difference, leads to higher initial superfluid acceleration as followed from Josephson-Anderson expression $m \dot{v}= \triangledown \mu$, i.e. the critical velocity is reached earlier in time. 
As mentioned in the main paper, the first generated vortex ring travels slower (with velocity extracted by a linear fit from $\vert x_{VR}\vert \geqslant 0.55 \simeq 2 w$) and has a lifetime which increases with increasing $z_0$ (Figs.~3(a) and 3(b) inset of main paper, respectively). As a result of this, the first VR for higher $z_0$ propagates further into the left reservoir (which has smaller condensate density) (Fig.~\ref{1stVR}(a)), maintaining a constant radius during this propagation for longer time than the smaller $z_0$ cases (Fig.~\ref{1stVR}(b)). \\
Moreover the first VR radius for $z_0=0.37$ during its propagation (flat area Fig.~\ref{1stVR}(b)) is larger than the one at $z_0=0.13$ which is consistent with its smaller propagation velocity value (shown in Fig. 3(a) main paper).

\subsubsection{Phase slippage}
In order to get more insight on the link between phase slippage and VR nucleation in our inhomogeneous BEC we 
show in Fig.~ \ref{slippage1} the time evolution of the relative phase $\Delta \phi _x$ along the $x$-direction 
for $V_0/\mu\simeq 0.8$ and $z_0=0.25$
at the location of the VR when first generated. Specifically in Fig.~ \ref{slippage1} this is
calculated for $z=0$ and $y \approx 0.5 l_x$, a value consistent with the location of the VR as seen by the density minima of the 2D and 3D density plots of Fig.~ \ref{slippage1}(a)-(b) respectively.
Fig.~ \ref{slippage1}(c) shows clearly that $\Delta \phi _x$ jumps locally by $\sim 2 \pi$ at a time $t=8.7$ms. 

\begin{figure}[h!] 
\begin{center}
\includegraphics[width=.97\columnwidth]{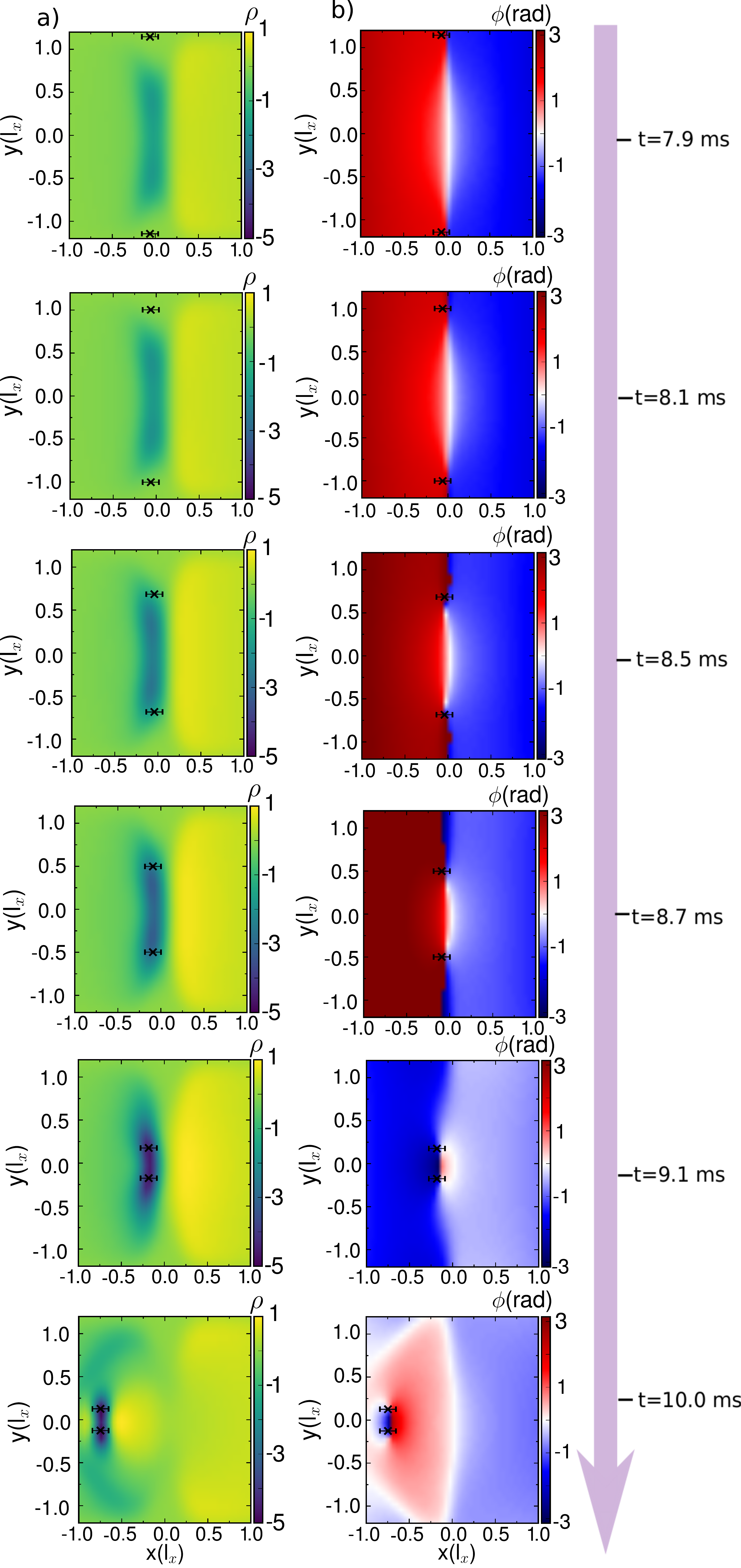}
\caption{Planar ($z=0$) 2D snapshots of (a) the BEC densities, and (b) the corresponding 2D phase profiles at different times, selected to cover the temporal window from the moment the VR enters the Thomas-Fermi surface at the barrier position, until when it leaves the barrier. 
The radii and positions of the tracked VR and their numerical error bars, associated with the uncertainty in the determination of the position due to the finite size of the vortex core, are denoted by `x' in the 2D phase profiles.
Same parameters as in Fig.~\ref{slippage1}.
}
\label{slippage}
\end{center}
\end{figure}

More details of this process are shown in Fig.~\ref{slippage} which plots the renormalized 2D density and phase profiles at different times during the initial VR dynamics,  and specifically from the time it is nucleated, until it enters the left reservoir. The initial phase jump of around 2$\pi$ at the vortex core position is indicated by an `x' at the 2D phase profile (see \ref{slippage}).
This picture is consistent with the phase slippage concept for superfluid helium. \cite{Anderson}.

\subsection{Backflow}

Here we provide more details about the interpretation of the drop in superfluid velocity reported in the main text in relation to Fig.~2(a).
\begin{figure}[h!]
\begin{center}
\includegraphics[width= \columnwidth]{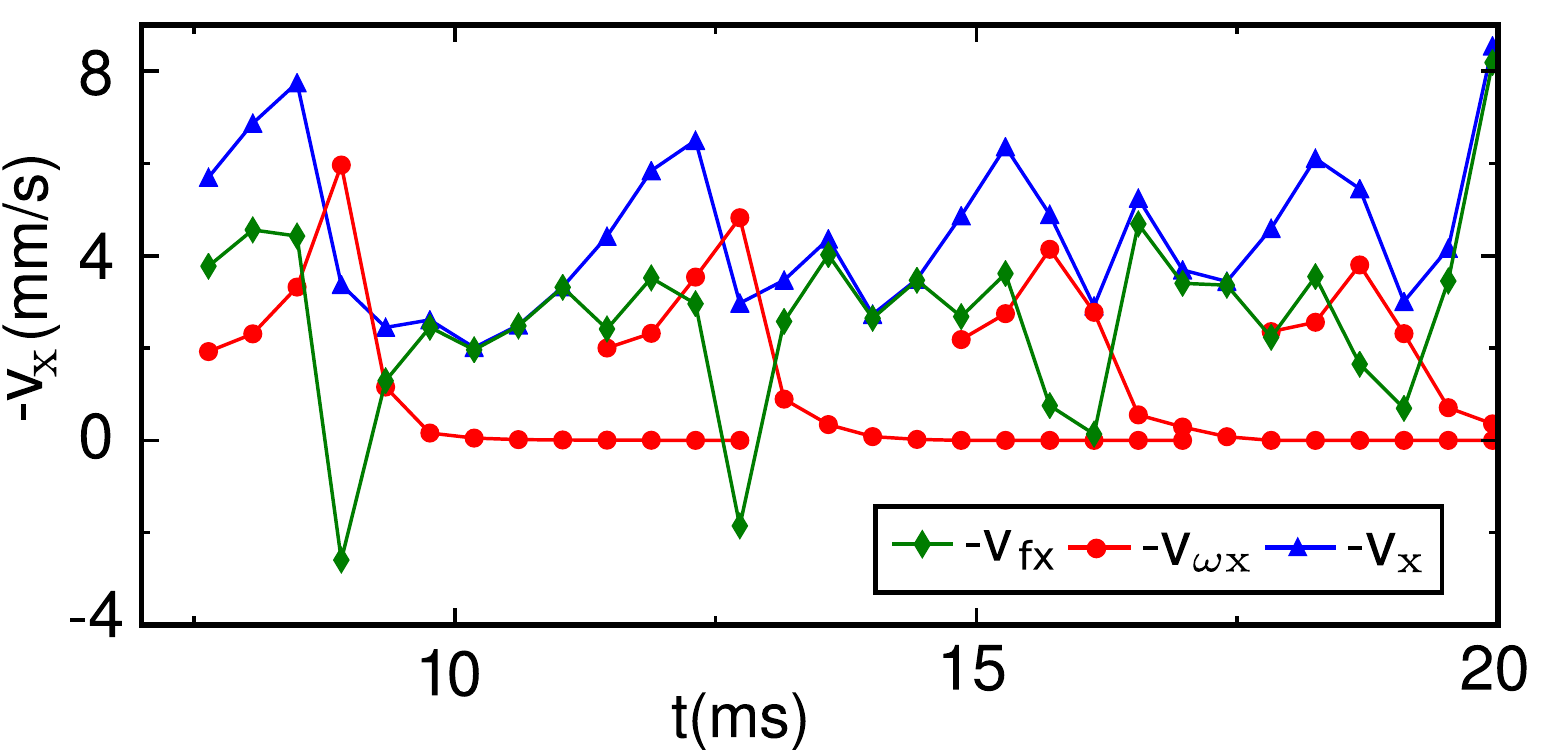}
\caption{Temporal evolution of the total superfluid axial velocity $v_x$, the vortex ring induced axial velocity $v_{\omega x}$ and the axial potential flow velocity $v_{fx}$ (called $v_f$ in the main text) in the center of the trap for the first four rings generated in the case $V_0/\mu\simeq 0.8$ and $z_0=0.25$.} 
\label{counterflow}
\end{center}
\end{figure}
As also noted in the main paper, we start by decomposing the superfluid velocity $\mathbf{v}$ into a superfluid potential flow $\mathbf{v}_{f}$ 
and in the flow generated by the superfluid singular vorticity distribution $\mathbf{v}_\omega$, via 
\begin{equation}
\mathbf{v} = \mathbf{v}_{f} + \mathbf{v}_\omega\;. 
\end{equation}
Given the complexity of the system studied,
as a first approximation we neglect the density gradient effects on the superfluid velocity. In addition, as the role of vortex images with respect to the BEC boundaries is still partially unresolved 
\cite{anglin-2002,mason,fetter-2009}, we do not consider the velocity field generated by the images of the VRs with respect to the boundaries of the condensate.

In order to calculate the superfluid potential flow $\mathbf{v}_{f}$ at each time $t$ in the center of the trap $O(0,0,0)$, we 
subtract the velocity field $\mathbf{v}_\omega(O,t)$ generated by the reconstructed VRs from
the total superfluid velocity $\mathbf{v}(O,t)$, {\it i.e.}

\begin{equation}
\mathbf{v}_{f}(O,t)=\mathbf{v}(O,t) -\mathbf{v}_\omega(O,t)\, \, ,\label{vs_flow}
\end{equation}
where $\mathbf{v}(O,t)$ is obtained via the phase gradient of the numerically computed wavefunction $\psi$ and $\mathbf{v}_\omega$ is calculated via the Biot-Savart integral \cite{hanninen}:
\begin{equation}
\displaystyle
\mathbf{v}_\omega(O,t) = -\frac{\kappa}{4\pi}\sum_{i=1}^N \oint_{_{\mathcal{C}_i(t)}}\!\!\frac{\mathbf{s}'(\zeta,t)\times \mathbf{s}(\zeta,t)}{|\mathbf{s}(\zeta,t)|^3}d\zeta  \, \, ,\label{vBS}
\end{equation}  
where $\kappa=h/m$ is the quantum of circulation, $N$ is the number of VRs present in the BEC, $\mathcal{C}_i(t)$ is the closed curve corresponding to the
$i$-th VR reconstructed via an algorithm based on the pseudo-vorticity vector \cite{rorai-etal-2016,villois-etal-2016}, $\mathbf{s}(\zeta,t)$ is the position
of the VR line-element corresponding to arclength $\zeta$ and $\mathbf{s}'(\zeta,t)$ its unit tangent vector. In Fig.~\ref{counterflow} we show the temporal evolution of the 
axial ($x$) components of $\mathbf{v}_{f}$, $\mathbf{v}_\omega$ and $\mathbf{v}$ in the centre of the trap $O$ for the parameters of Fig.~2 of the main paper ($V_0/\mu \simeq 0.8$ and $z_0=0.25$).

We observe that in the event of a vortex ring generation, the superfluid flow  $v_{fx}$ slows down and even reverses its sign for the first two rings generated: 
the nucleation of vortex rings leads to a reduction in the main flow, thus slowing down the population imbalance dynamics. 

In fact, the corresponding reduction in $I$ is clearly visible in both Figs.~\ref{vx} and \ref{I_z025}, and it is particularly pronounced for the only (Fig.~\ref{vx}) or the first (Fig.~\ref{I_z025}) generated VR, for which it even reverses sign (i.e.~$-I < 0$).


\section{Vortex Ring Energy calculation}
The total energy of the system can be decomposed as following:
\begin{equation}
E=\int \left[ \displaystyle e _{int} (\vec{r},t)+ \displaystyle e_k (\vec{r},t) + \displaystyle e _{V} (\vec{r},r) \right] d^3 \vec{r}
\end{equation}
where 
\begin{equation}
e _{int} (\vec{r},t)= \frac{g}{2} \rho ^2
\end{equation} 
is the interaction energy density, 
\begin{equation}
e _{V}  (\vec{r},t)= \rho V_{trap}
\end{equation}
is  the potential energy,  and
\begin{align}
\displaystyle e_k (\vec{r},t)  =\frac{1}{2}\vert \nabla \psi\vert ^2 &= \frac{1}{2} \rho \mathbf{v} ^2+\frac{1}{2} \vert  \nabla \sqrt{\rho(\vec{r},t)}\vert ^2  \\
  & =\frac{1}{2} \rho \mathbf{v} ^2+\displaystyle e_q (\vec{r},t)
\end{align}
is the kinetic energy density where  $\displaystyle e_q (\vec{r},t)$ is the quantum pressure term. The quadratic term involving the velocity of the flow can be split into incompressible $\displaystyle e_k ^i(\vec{r},t)$ and compressible $\displaystyle e_k ^c(\vec{r},t)$ contributions as follows: 
\begin{equation}
\frac{1}{2} \rho \mathbf{v} ^2= \frac{1}{2}\left [ \left ( \sqrt{\rho} \mathbf{v} \right )^i \right ]^2+\frac{1}{2}\left [ \left ( \sqrt{\rho} \mathbf{v} \right )^c \right ]^2=\displaystyle e_k ^i(\vec{r},t)+\displaystyle e_k ^c(\vec{r},t)
\end{equation} 
where  $\vec{\nabla} \cdot (\sqrt{\rho}\vec{v}) ^i=0$ and  $\vec{\nabla}$x$(\sqrt{\rho}\vec{v}) ^c=0$. The fields $(\sqrt{\rho}\vec{v}) ^i$ and $(\sqrt{\rho}\vec{v}) ^c$  are calculated employing the Helmholtz decomposition \cite{nore1997, numasato,horng, griffin2019}.

\begin{figure}[t!]
\centering
  \includegraphics[width=.9\linewidth]{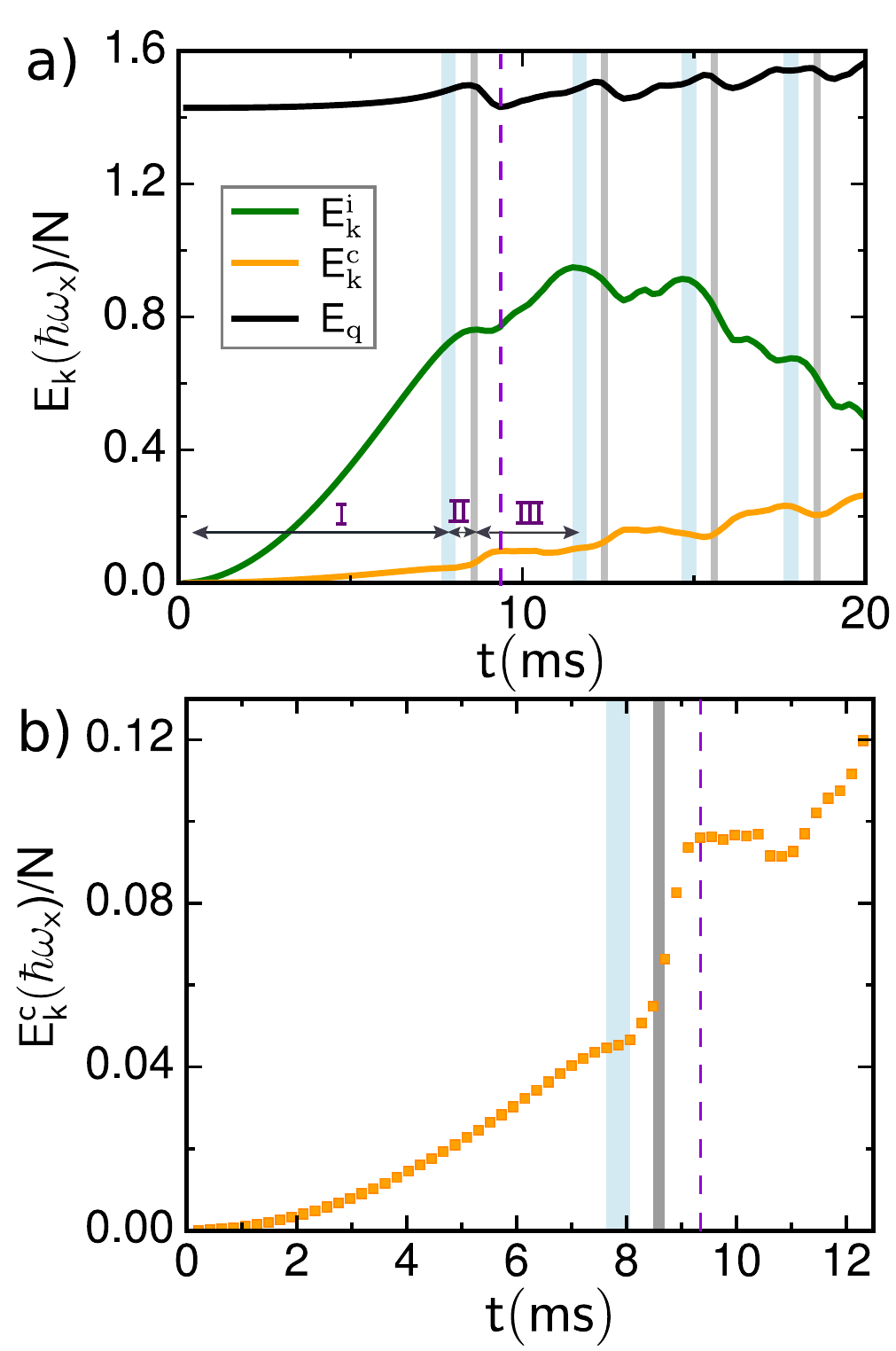}
  \caption{a) The  temporal evolution of $ E_k ^i$, $ E_k ^c$ and $E_q$. Showing also the three stages of the VR nucleation discussed in the main text. b) The temporal evolution of  $ E_k ^c$ till the nucleation of the second VR. As in Fig. 2 of the main paper, vertical blue shaded areas in both subplots highlight the moment when each VR is nucleated and the grey shaded areas the moment when each VR enters the local Thomas-Fermi surface (already shown in Fig.~2 of the main paper). Vertical dashed line indicates the moment during the axial propagation of the VR, when it reaches the point of largest transversal spatial extent corresponding to the edge of the barrier, after which time the background density begins decreasing very slowly due to the underlying external harmonic trap along x-axes. \\}
  \label{fig:EKi}
\end{figure}
In Fig. 2(d) of the main paper we have shown the temporal evolution of the integrals of $\displaystyle e_k ^i(\vec{r},t)$ and $\displaystyle e_k ^c(\vec{r},t)$ (called $E_k ^i$ and $E_k ^c$ respectively). In Fig. \ref{fig:EKi} a) instead we show also the integral of $\displaystyle e_q(\vec{r},t)$ (called $E_q$) which is larger than $E_k ^i$ and  $E_k ^c$ and it is slightly affected by the VR nucleation. As discussed in the main paper, when each VR enters the TF surface (vertical grey shadow areas in Fig. \ref{fig:EKi}, shown also in Fig. 2 of the main paper), we see an increase of the compressible energy. This step-like increase is shown better in Fig. \ref{fig:EKi} (b) where only $E_k ^c$ is shown until the second VR is nucleated. The step-like increase starts when the first VR enters the Thomas-Fermi surface and stops (reaches a 'plateau') when the VR is at the position corresponding to maximum transverse Thomas-Fermi spatial extent, which for this case occurs around t$\sim$ 9.3 ms (shown by vertical dashed line in figure \ref{fig:EKi}). Moreover, the ratio of the increase of $E_k ^c$ to the compressible energy before the first VR nucleation takes the values $1.09\, , 1.36\, , 1.57$ for the nucleation of the $1^{\rm st}$,$ 2^{\rm nd}$, $3^{\rm rd}$ VRs, respectively. 

\begin{figure}[h!]
\begin{center}
\includegraphics[width= 0.99\columnwidth]{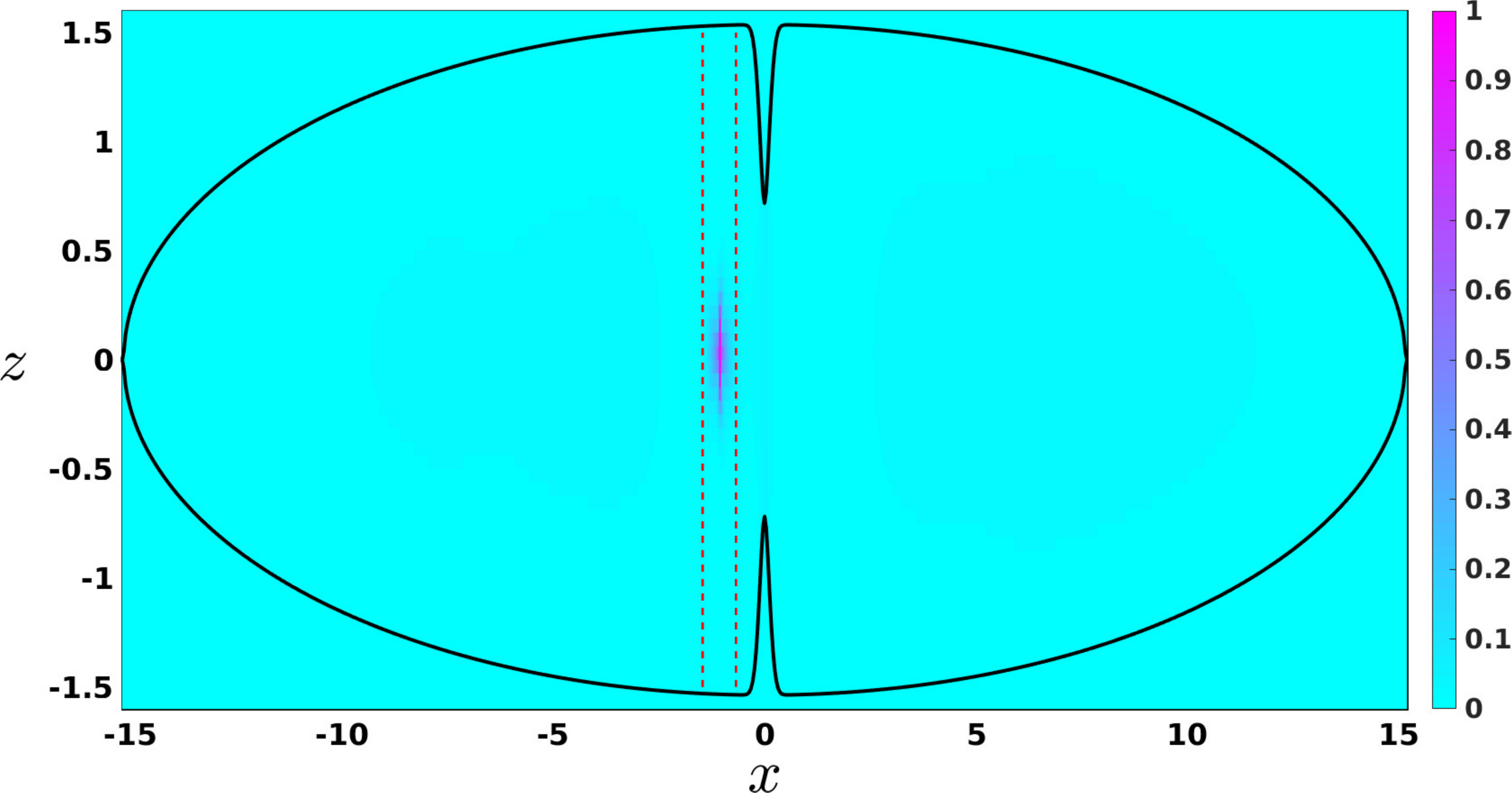}
\caption{Two-dimensional plot of the incompressible kinetic energy density per unit mass $e_k^i$ in units of $\hbar\omega_x/(M\ell_x^3)$. 
The volume between the red dotted lines is the region $\mathcal{R}$.} 
\label{e_kin_density}
\end{center}
\end{figure}

In Fig.~3(a) of the main paper, we have instead used the incompressible kinetic energy of the generated VRs (shown as a function of $z_0$ ) to further characterize their motion.
The incompressible kinetic energy of the VR $E_{k,VR}^i$, which depends on $v_\omega$ only, is obtained by the following procedure. 
We first integrate the incompressible 
kinetic energy density per unit mass
$\displaystyle e_k^i$ 
 on the volume $\mathcal{R}$ encompassing the vortex ring.
The region $\mathcal{R}$ is defined as follows
\begin{eqnarray}
\displaystyle
R=&& \{ (x,y,z) : x_{V\!R}-\Delta x<x<x_{V\!R}+\Delta x \;\; ; \nonumber \\[1mm] 
  &&   \;\;\;\;\;\;\;\;\;\;\;\;\;\;\;\; -R^{TF}_y(x)<~y~<R^{TF}_y(x) \;\;\;\;\; ; \nonumber \\[1mm]
  &&   \;\;\;\;\;\;\;\;\;\;\;\;\;\;\;\; -R^{TF}_z(x)< ~z~<R^{TF}_z(x) \;\;\} \;\; , \nonumber
\end{eqnarray}
where $x_{V\!R}=-l_x$ and $\Delta x=4R_{VR}$, where $R_{VR}$ is the VR radius when $x_{V\!R}=-l_x$. 
(For improved clarity, a visualization of the region $\mathcal{R}$ is given in Fig.~\ref{e_kin_density}). \\

\begin{figure*}[htb!] 
\centering
 \makebox[\textwidth]{\includegraphics[width=.8\paperwidth]{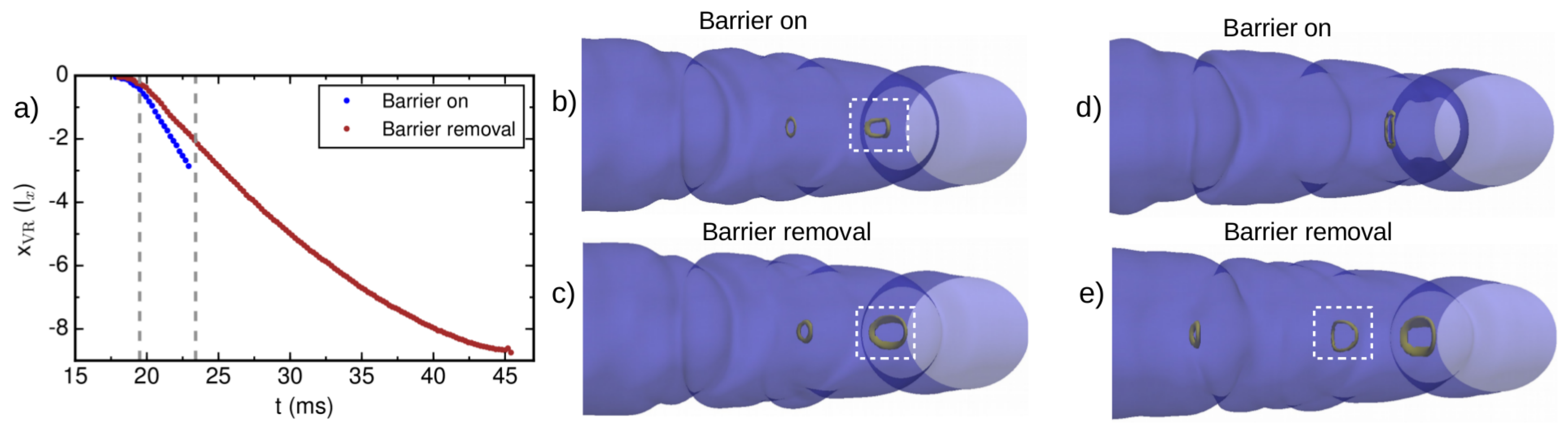}}
\caption{(a) Time evolution of x-position of the 4th VR [shown in the white box] for both barrier on and barrier removal for initial $V_0/\mu = 0.8$ and $z_0 =0.25$ (discussed in main paper Fig.~4). Density isosurface (taken at 5$\%$ of maximum density) for (b) barrier on, and (c) barrier removal at t=19.5 ms. (d)-(e) Corresponding profiles at t=23.4 ms.
}
\label{quench08munew}
\end{figure*}
The values of $x_\mathrm{VR}$ and $\Delta x$ are chosen stemming from the following considerations.
First, in a homogeneous and unbounded BEC, a cylindrical region, coaxial with respect to a VR, of height $8R_{VR}$ (the VR center being 
placed in the center at $4R_{VR}$) and radius $5R_{VR}$ contains more than $90\%$ of the incompressible kinetic energy of the VR.
Second, the flow velocity is negligible  in $\mathcal{R}$, {\it i.e.} $\mathbf{v}_f\approx 0$ and hence $\mathbf{v}\approx \mathbf{v}_\omega$.
Therefore, the incompressible kinetic energy present in $\mathcal{R}$ only originates from the VR.  
As a consequence, $\displaystyle E_{k,VR}^i=\int_{\mathcal{R}}e_k^id\mathbf{x}$
is a reasonable estimate of the VR incompressible kinetic energy. Moreover, the VR is still close to its nucleation region and
therefore $E_{k,VR}^i$ can, at least qualitatively, be considered as proxy for the VR initial incompressible kinetic energy. 
Investigating the dependence of $E_{k,VR}^i$ on the initial population imbalance $z_0$, shown in Fig.~3(a) of the main paper, 
we find that $E_{k,VR}^i$ is an increasing function of $z_0$.

To assess the impact of confinement and inhomogeneity, we also evaluate the corresponding energies of the first nucleated VRs if they had been immersed
in an unbounded and homogeneous BEC employing the analytical formula derived by Roberts and Grant \cite{roberts1971}:
\begin{equation}
\displaystyle
E_{k,VR}^{i,RG}=\frac{1}{2}\, \rho \, \kappa^2 R_{VR} \left [ \ln \left ( \frac{8R_{VR}}{\xi}  \right ) - 1.615\right ]\,\, ,
\end{equation}   
where $\kappa$ is the quantum of circulation and both the density $\rho$ and the healing length $\xi$ have been evaluated on the $x$-axis for $x\sim x_{VR}$
(allowing $\rho$ to regain its bulk value not perturbed by the presence of the VR). The result of this calculation is reported in Fig.~3(a) of the
main manuscript (pink triangles) showing that confinement and inhomogeneity play a fundmamental role in the energetics of the system studied \cite{groszek2018,donadello2014}.

\section{Dissipation of kinetic energy}
In the main manuscript, to quantify the irreversible dissipation of Josephson oscillations, we introduce the acoustic dissipation $\epsilon_c$ and the 
incompressible dissipation $\epsilon_i$. The definition of these quantities is as follows:
\begin{eqnarray}
\displaystyle
\epsilon_c & = & \Delta E_k^c/E_{k,[0,1]}^{tot} \\[3mm]
\epsilon_i & = & E_{k,VR}^i/E_{k,[0,1]}^{tot}  
\end{eqnarray}
\noindent
where $E_{k,[0,1]}^{tot}$ is the total kinetic energy flowing through the junction until the nucleation of the first VR, $E_{k,VR}^i$ is the incompressible 
kinetic energy of the first VR (as illustrated in the previous section) and $\Delta E_k^c$ is the increase of compressible energy in the system observed 
in correspondence of the nucleation of the first VR ([Fig. 2(d)] in the main paper).

\section{Kelvin waves and barrier removal}

Fig.~4(a)-(b) of the main paper discussed how the barrier removal process -- implemented experimentally prior to time-of-flight observation -- affects the VR dynamics.
Specifically, we first let the system evolve for 13 ms, and subsequently we remove the barrier linearly over a period of 40 ms, and observe the motion of the fourth generated VR (i.e.\ VR generated around $t \sim 17.5$ ms in Fig.~2 of main paper). 
As commented in the paper, this process significantly extends the lifetime of this VR. 

More details of this effect are given in  Fig.~\ref{quench08munew}, which compares the evolution of the x-position of the 4th VR in the case of barrier on (blue points in Fig.~\ref{quench08munew}(a)) and barrier removal (red points), showing also a direct comparison of appropriate 3D density isosurfaces which reveal the vortex rings in both cases at times $t \approx 19.5$ ms (Fig.~\ref{quench08munew}(b)-(c)) and $t \approx 23.4$ ms (Fig.~\ref{quench08munew}(d)-(e)).
From the slope of  Fig.~\ref{quench08munew}(a), we deduce that the VR velocity in the case of barrier removal is smaller than the one with barrier on. This is consistent with a larger propagating VR radius in the case of barrier removal, an effect visible by comparing the size of the VR highlighted inside the white box in the 3D density plots of Fig.~\ref{quench08munew}(b)-(c).
Specifically at $t=19.5$ms, the image shows two VRs (respectively the third and fourth VRs generated), 
while a few ms later, at 23.4ms, we see clearly that only one VR remains in the case of barrier on (Fig.~\ref{quench08munew}(d)), whereas three VRs are still visible in the corresponding case of gradual barrier removal (Fig.~\ref{quench08munew}(e)).
Note that, in the case of barrier removal, the total number of VRs generated is less than the corresponding case with barrier on, because the decrease in the barrier height leads to an increase in the density inside the barrier, thus increasing the local speed of sound, which in turn decreases the superfluid velocity to below the speed of sound $c=\sqrt{g n/M}$ -- a process which prohibits further VR generation.

\begin{figure*}[t!]
\begin{center}
\includegraphics[width= 0.6\textwidth]{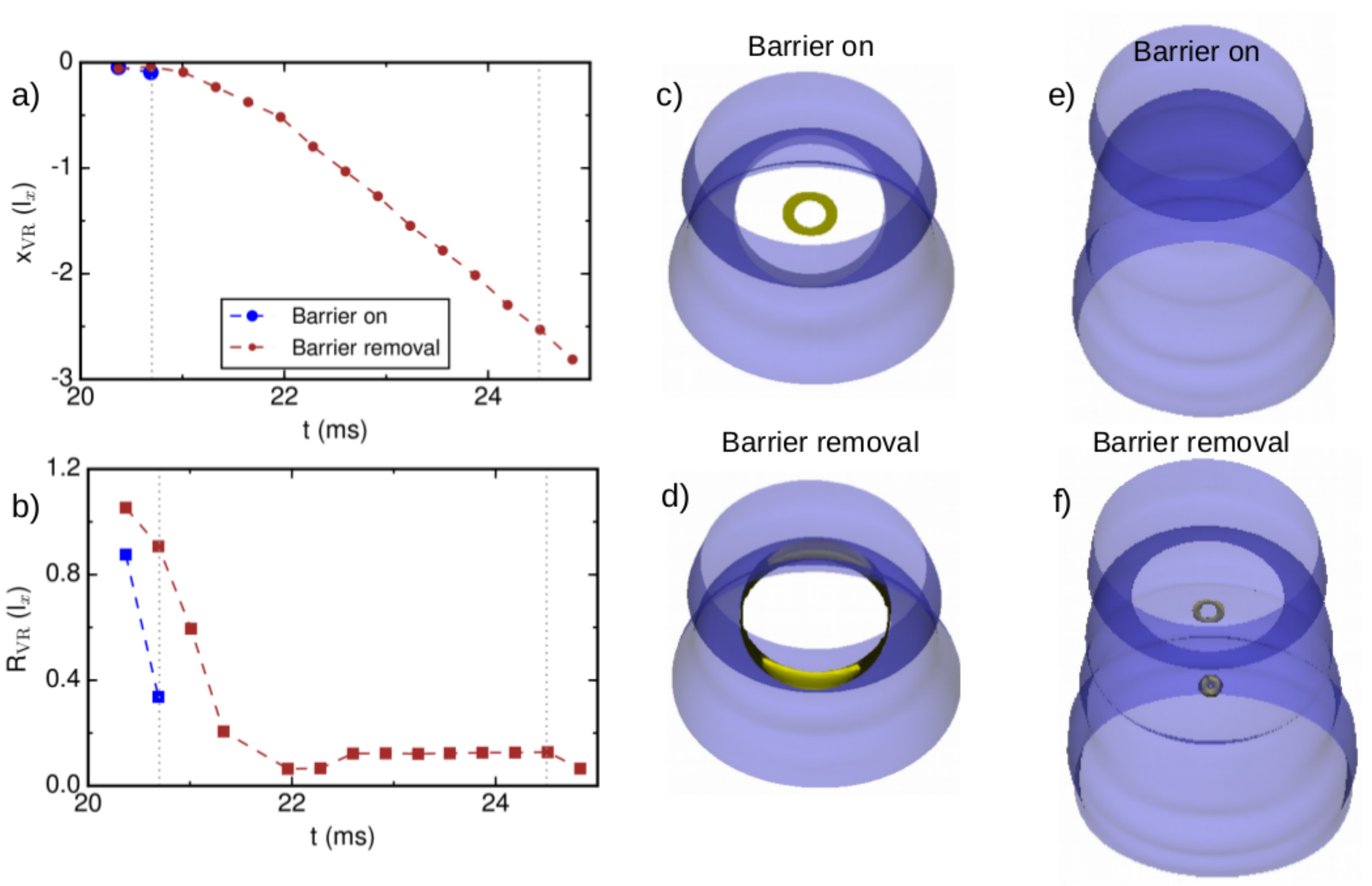}
\caption{(a)-(b) Comparison of (a) $x$-position, and (b) VR radius in the case when barrier height is kept fixed (blue points), or gradually decreased (red points) for  $V_0/\mu =1.2$ and $z_0 = 0.25$. 
Density isosurface (taken at 5$\%$ of maximum density) at (c)-(d) $t$=20.7ms  and (e)-(f) 24.5ms for both cases of barrier on or removed.
} 
\label{quench}
\end{center}
\end{figure*}


Another crucial point to note in relating our findings to experimental observables is that in the experiment \cite{Lidiss} they observe vortices propagating in the system even for initial barrier height $V_0 > \mu$. In our simulations with fixed barrier height $V_0 > \mu$, the VRs generated inside the barrier shrink fast without entering in the bulk,
whereas the experimentally relevant barrier removal enables their detection in the bulk. This is shown in Fig.~\ref{quench} for $V_0/\mu=1.2$ for both cases of barrier on and barrier removal. 

Specifically, we observe that once the barrier height, while decreasing its value, reaches some characteristic value (here $\sim 0.9 \mu$) the VR is able to escape the barrier region and propagate in the left reservoir (red points in Fig.~\ref{quench}(a)), unlike the corresponding case of constant barrier height $V_0 > \mu$ (blue points). 
As visible in the density plots at t=20.7 ms (Fig.~\ref{quench}(c)-(d)),  the VR in the case of barrier on (Fig.~\ref{quench}(c)) has a much smaller radius that the corresponding one when the barrier is removed (Fig.~\ref{quench}(d)), i.e. a much smaller energy. For this reason at the subsequent t=24.5 ms, the VR with barrier on (Fig.~\ref{quench}(e)) has already shrunk while the one generated under gradual barrier removal (Fig.~\ref{quench}(f)) propagates inside the superfluid. In fact the removal can in this case (t=24.5 ms) facilitate the simultaneous observation of two VRs, in stark contrast to the barrier on case which reveals none. 

\subsection{Kelvin Waves}
\begin{figure}[h!]
\begin{center}
\includegraphics[width= 0.9\columnwidth]{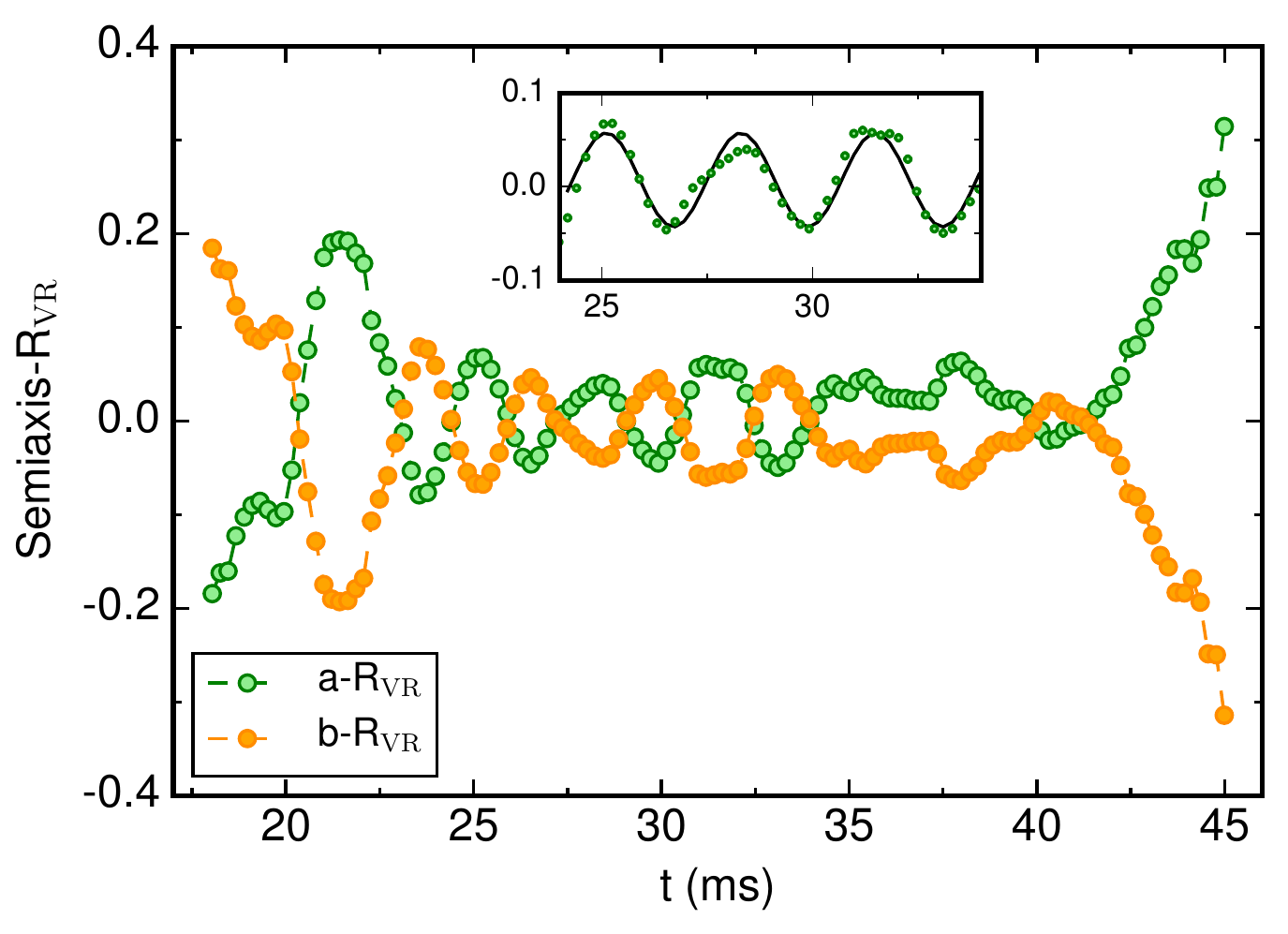}
\caption{Evolution of deviations $(a-R_{VR})$ (green points) and $(b-R_{VR})$ of the VR semiaxes from their average value. Inset: sinusoidal fit (black line) to $(a-R_{VR})$ in the time interval [24,34]ms, confirming the KW nature of the VR excitations. } 
\label{KW}
\end{center}
\end{figure}

The main paper discussed the role of Kelvin Wave (KW) excitation on the VR dynamics, already visible in Figs.~4(a)-(b) of main manuscript.  Here we provide further details on this characterization. 
Due to the anisotropy in the transverse direction ($\omega _y \neq \omega _z$) the VR shape is elliptic when it is nucleated. During its propagation the VR shape oscillates by inverting its elliptical semiaxis, representing an $m=2$ KW excitation of the circular shape. In fact, the VR 2D profile is best fit by the function $(y/a)^2+(z/b)^2=1$. If we define $R_{VR}=(a+b)/2$, the deformation of the elliptic VR from its ideal circular shape with radius $R_{VR}$ is $(a-R_{VR})$ along the y-direction and $(b-R_{VR})$ along the z-direction. Fig.~\ref{KW} shows the time evolution of these quantities in the case of the 4th VR undergoing the barrier removal process over the relevant post-generation temporal window $t \in [17.5, \, 45]$ ms. After such time the VR is destroyed by interaction with the condensate boundary, an effect already studied in different contexts in \cite{anglin-2002, mason, Levin, Mateo}.

In the limit of a VR radius much larger than the core size, the period of oscillations of KW is given by the dispersion relation $\displaystyle \omega (k)\sim \kappa k^2/(4\pi)[ \ln (2/(k\xi)-0.5772 ]$ \cite{barenghi-etal-1985}, with $k$ the wavenumber, $\xi$ the vortex core and $\kappa =h/M$ the quantum of circulation. The KW wavenumber $k$ is found from the wavelength $\lambda =2 \pi / k$, which satisfies the relation $\lambda m =2 \pi R$ for $m=2$ and $R$ the VR radius.  Estimating the vortex core inside the bulk as being comparable to the molecular BEC coherence length ($\xi\simeq 0.5 \mu$m) and approximating $R$ as the mean value of the 4th VR radius in the time interval [24, 34] ms, the above dispersion relation predicts a KW period $\tau \simeq$ 3.3 ms,which is found to be in excellent agreement with a sinusoidal fit to our numerically extracted values of $(a-R_{VR})$ in the range [24, 34] ms (black line in inset of Fig. \ref{KW}) which yields $\tau =(3.19\pm0.03)$ ms; we note that this agreement is excellent, even though the VR radius is only five times the vortex core. 

\section{T$>$0 Vortex Ring Dynamics}
The main paper has shown that, in the range $0 \le T \le 0.4 T_{\rm c}$ considered (for which $0.4 V_0 < k_B T < 0.8 V_0$) finite temperatures have practically no effect on the VR generation process, and the subsequent early VR dynamics, provided the condensate number is fixed to the corresponding $T=0$ value.
In this case, the main effect is the addition of an extra potential to the BEC (Eq.~(\ref{gped})-(\ref{gped2})), as previously observed in the context of vortex dynamics and reconnections \cite{vortexT1,vortexT2,vortexT3}. This is because such timescales are typically much shorter than corresponding ones for dissipation due to the relative motion between the BEC and the thermal cloud .
At the same time, temperature has a notable cumulative effect on the overall post-generation VR dynamics: specifically, it reduces the VR lifetime (shown in Fig.~3(b) (inset) of the main paper for the first generated VR and different $z_0$) and breaks its motional symmetry (main paper Fig.~4(c)). This effect was already shown to become more pronounced in the cases of barrier removal, because in such cases the VR lifetimes are considerably longer. 
Here we provide more details on the effect of the thermal cloud on the VR dynamics.
\begin{figure}[h!]
\begin{center}
\includegraphics[width= \columnwidth]{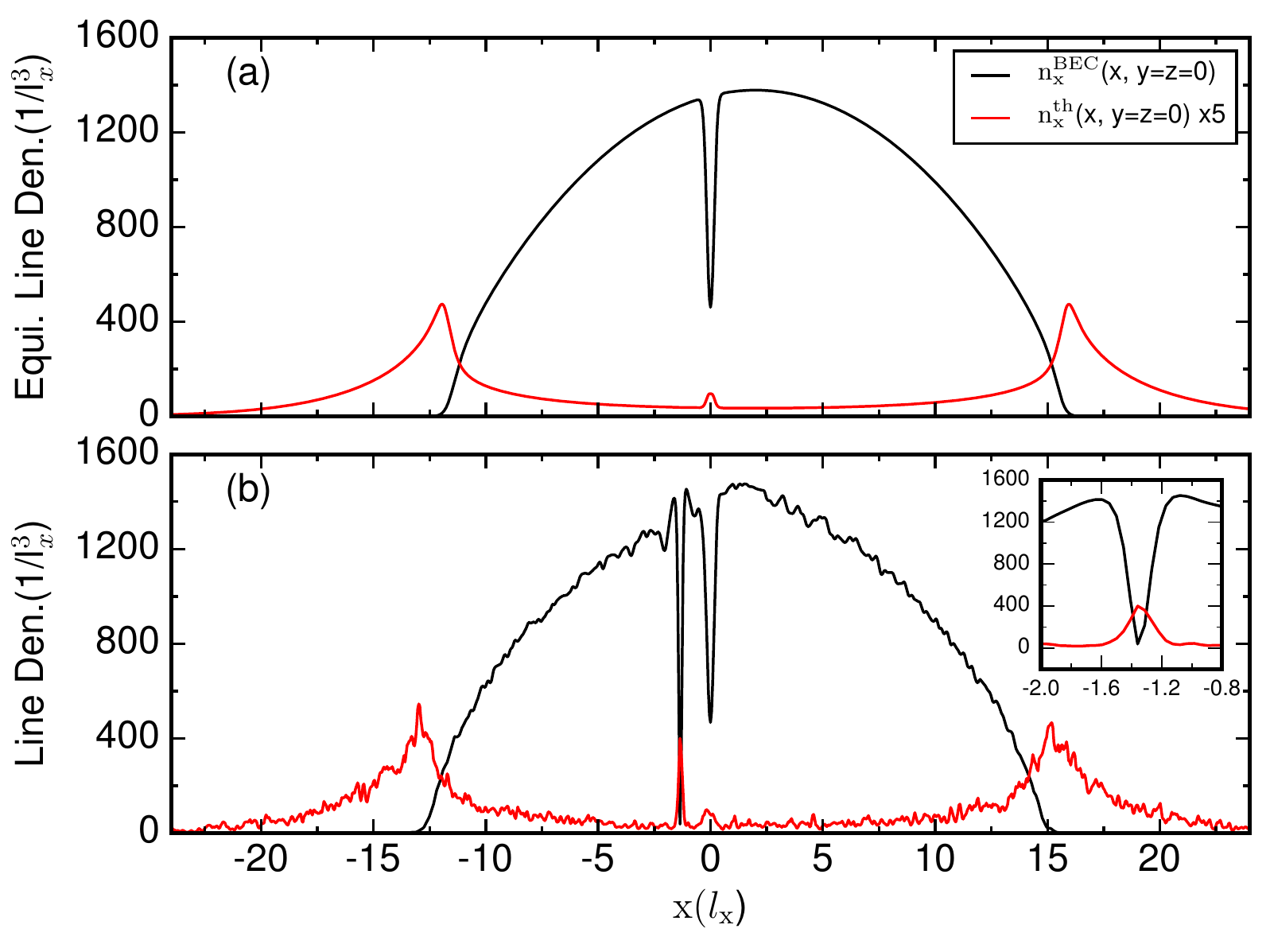}
\caption{Density profile along x-direction (y=z=0) of the condensate (black line) and thermal cloud (red line) at equilibrium (a) and at 10.6 ms time evolution (b) where a VR is present (at $x \sim -1.4 l_x$) for $V_0/\mu \simeq 0.8$, $z_0=0.25$ and T$\simeq 0.4 T_c$. The inset in (b) shows the condensate and thermal density around the VR position.} 
\label{denT}
\end{center}
\end{figure}

For easier visualization, Fig.~\ref{denT} shows an example of the axial condensate (black) and thermal cloud density (red) for  $V_0/\mu \simeq 0.8$, $z_0=0.25$ and T$\simeq 0.4 T_c$.  In this figure, density is plotted along the x-direction (for $y=z=0$) both (a) at equilibrium, and (b) at a later time ($t=10.6$ ms) when the first VR has already entered the left well (visible for $x \sim -1.4 l_x$). 
As expected, the thermal cloud has local maxima at both the barrier position, and at the edges of the condensate where the BEC density has local minima.
This is because of the repulsive interaction between the thermal cloud and the condensate. Moreover when a VR is present, its core is filled by thermal molecules, as seen clearly in Fig.~\ref{denT}(b) (around $x \sim -1.4 l_x$). This effect has already been reported in \cite{vortexT1,vortexT2,vortexT3}. 

\begin{figure}[h!]
\begin{center}
\includegraphics[width= 0.9\columnwidth]{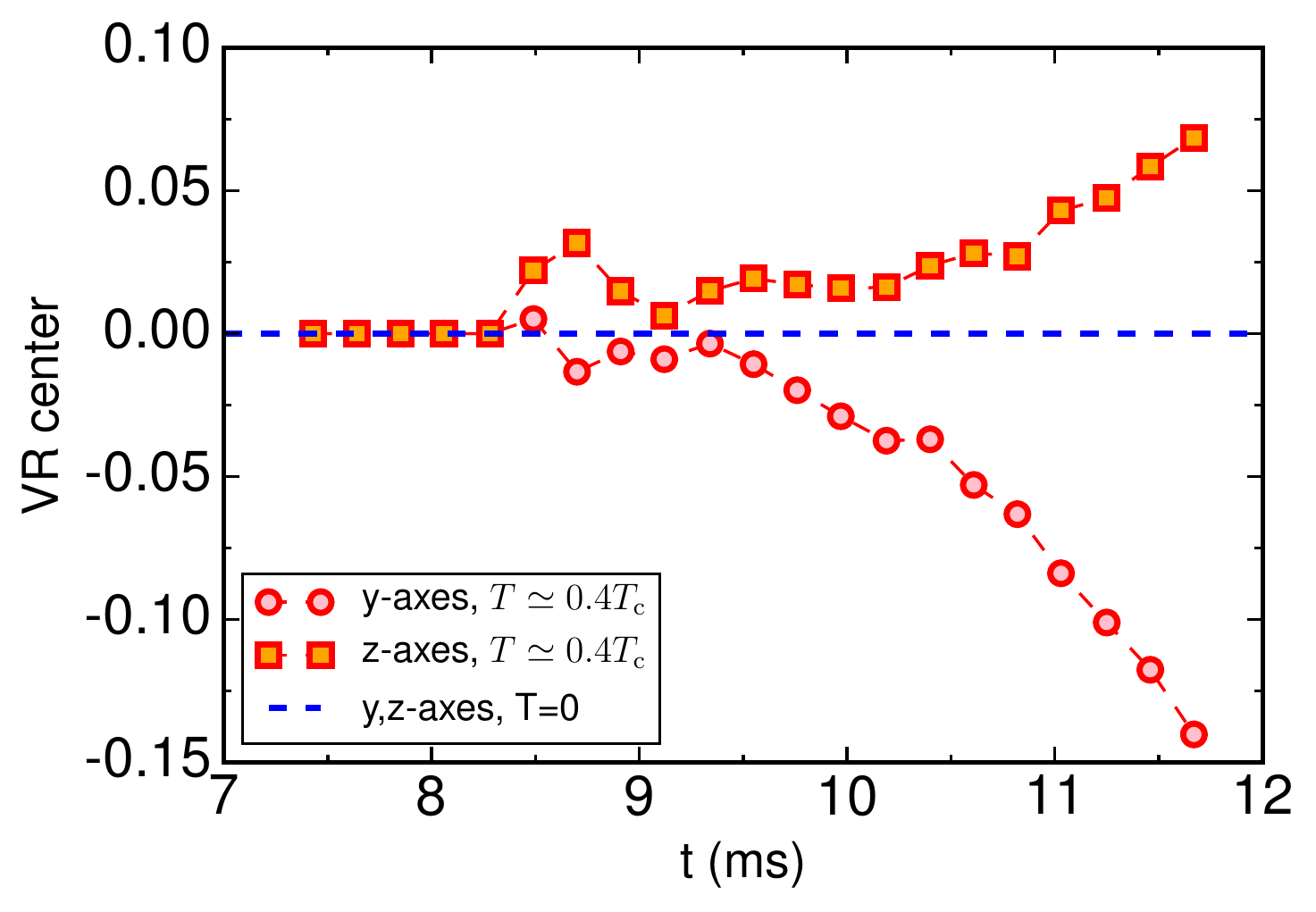}
\caption{The first VR center along the $y$ and $z$ directions at $T=0$ (horizontal dashed line) and at T$\simeq 0.4 T_c$ (red symbols) in the case of barrier removal with initial $V_0/\mu \simeq 0.8$ and $z_0=0.25$.} 
\label{offcenter}
\end{center}
\end{figure}

\begin{figure*}
\begin{center}
\includegraphics[width=0.8\linewidth]{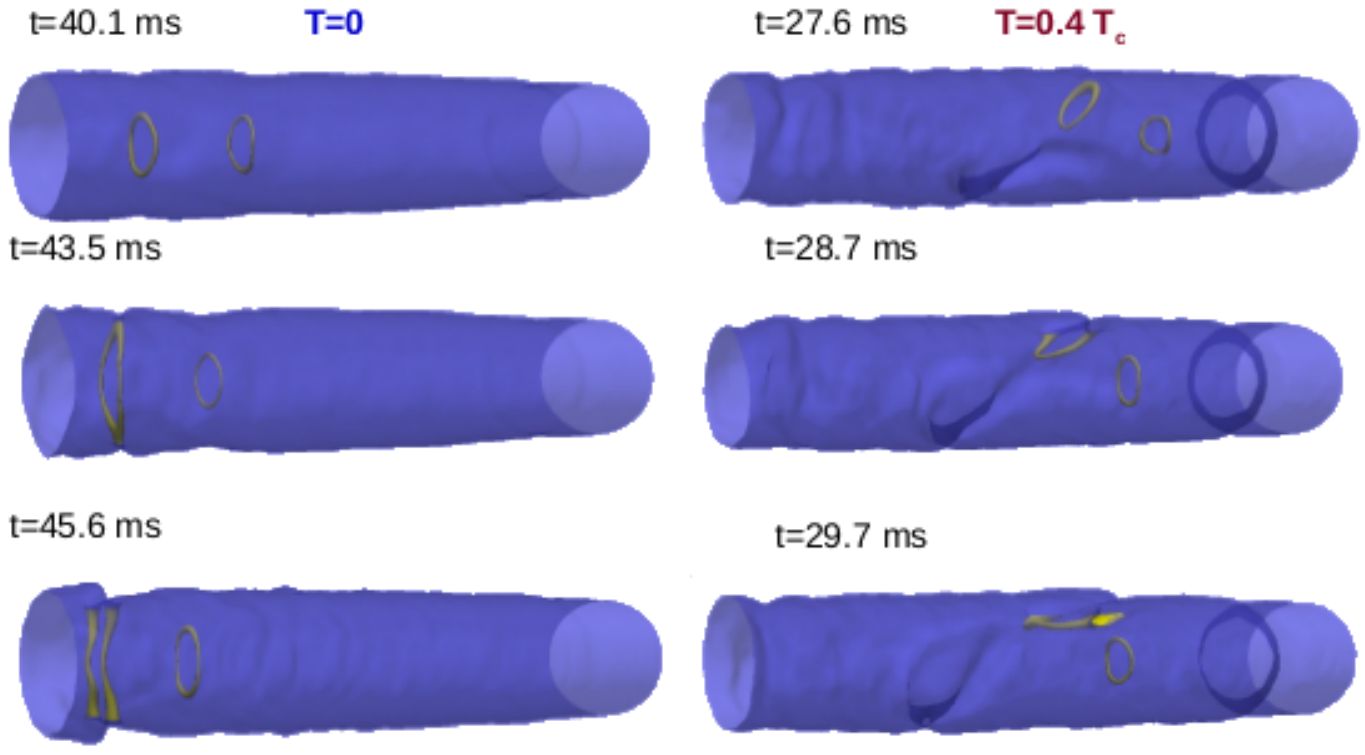}
\caption{Density isosurface (taken at 5$\%$ of maximum density) for the parameters of Fig.~\ref{offcenter}.
The fourth VR discussed in the main paper is the leftmost VR in each plot (the other visibile VR is the subsequently-generated fifth VR).
} 
\label{offT}
\end{center}
\end{figure*}
Even though the considered temperature range (with $k_B T < 0.8 V_0$) has a small effect on the VR generation process, nonetheless it does exert a `drift force' causing the VR to go off-center while propagating along the negative $x$-direction. 
A clear visualization of this effect is shown in Fig.~\ref{offcenter}, showing the time evolution of the fourth VR center along the $y$ and $z$ directions 
for a characteristic single numerical realisation when the barrier is gradually removed (corresponding to Fig.~4(c) of the main paper) . 
Corresponding 3D densities are shown in Fig.~\ref{offT}, clearly contrasting the $T=0$ to the $T>0$ case, providing an alternative visualization to that of Fig.~4(b)-(c) of the main paper.
Note that while the lifetime is well predicted within our kinetic model, the precise details of the VR trajectory -- i.e. exactly how it goes off center and approaches the boundary -- are sensitive to the numerical realisation.

\clearpage
\end{document}